\documentclass{article}
\usepackage{latexsym}
\usepackage{graphicx}
\usepackage{epsfig}
\newcommand \be {\begin{equation}}
\newcommand \bea {\begin{eqnarray}}
\newcommand \ee {\end{equation}}
\newcommand \eea {\end{eqnarray}}

\newcommand{\bit}{\begin{itemize}}
\newcommand{\eit}{\end{itemize}}
\newcommand{\Z}{\mathbf{Z}}

\newcommand{\C}{\mathbf{C}}

\newcommand{\eps}{\epsilon}
\newcommand{\beps}{\bar\epsilon}
\newcommand{\bpsi}{\bar\psi}
\newcommand{\blambda}{\bar\lambda}
\newcommand{\bsigma}{\bar\sigma}
\newcommand{\dalpha}{\dot\alpha}
\newcommand{\dbeta}{\dot\beta}

\newcommand{\lra}{\leftrightarrow}

\newcommand{\ra}{\rightarrow}
\newcommand \dsl {\not\!\partial}

\begin{document}
\topskip 2cm

\begin{titlepage}

\rightline{UNIL-IPT-00-15}
\vspace{10mm}

\begin{center}
{\Large\bf Supersymmetric Noether Currents} \\ 
\vspace{.2cm}
{\Large\bf and Seiberg-Witten Theory} \\
\vspace{2.5cm}
{\large A. Iorio$^{* \dag}$, L. O'Raifeartaigh$^{\dag}$, and S. Wolf$^{+}$} \\
\vspace{.5cm}
{\sl $^{*}$ Department of Physics, University of Salerno,}\\
{\sl Via Allende, Baronissi (SA), Italy} \\
{\sl $^{\dag}$ School of Theoretical Physics,}
{\sl Dublin Institute for Advanced Studies}\\
{\sl 10 Burlington Road, Dublin 4, Ireland} \\
{\sl $^{+}$ Institut de Physique Th\'eorique, BSP}\\
{\sl  1015 Lausanne, Switzerland} \\
\vspace {.2cm}
{\sl e-mail: iorio@maths.tcd.ie ; lor@stp.dias.ie ; sylvain.wolf@ipt.unil.ch}\\
\vspace{.5cm}

\vspace{2.5cm}

\begin{abstract}
\noindent
The purpose of this paper is twofold. The first purpose is to review a 
systematic construction of Noether currents for supersymmetric theories, 
especially effective supersymmetric theories. The second purpose is to 
 use these currents to derive the mass-formula for the quantized 
Seiberg-Witten model from the supersymmetric algebra. We check that the 
mass-formula of the low-energy theory agrees with that of the full 
theory (in the broken phase).  \end{abstract}
\end{center}

\vfill

\end{titlepage}

\newpage

\section{Introduction}

\noindent 
Most aspects of Supersymmetry (Susy) have by now been extensively 
developed. However, supersymmetric Noether currents do not appear to 
have been treated in a systematic manner in the literature. This is 
a pity because many aspects of Susy could be clarified by considering 
the currents, an example being the computation of the mass formula 
in the Seiberg-Witten (SW) model \cite{sw} (see also \cite{bil} for a review).
The Susy Noether currents correspond to space-time symmetries and hence are by
no means trivial, as discussed in more detail below. 
Furthermore, in models, such as the SW model, the Susy theories are effective 
ones, and in effective theories the Noether currents have particular 
difficulties, as is also discussed below. So far, there does not seem to be 
any systematic approach to constructing the currents. In two previous 
letters \cite{ior}, \cite{wolf},
an attempt was made to remedy this situation, using the SW model 
as a non-trivial test case. The purpose of the present paper is to 
give a more detailed review of this more systematic construction of 
Susy Noether currents and to apply it to the SW-model. 

\noindent
From the point of view of the Noether currents the essential feature 
of the SW model is the existence of a central charge $Z$, which is 
obtained as an anti-commutator of the Susy charges and provides the 
mass formula. Thus the Noether currents are directly connected to 
the mass-spectrum. The mass formula is given by 
\be\label{mass}
M = |Z| = \sqrt2 |n_e a + n_m a_D|
\ee
and it is of paramount importance for the exact solution of the quantum 
theory. 
In Eq. (\ref{mass}) $n_e$ and $n_m$ are the electric and magnetic 
quantum numbers respectively, and $a$ and $a_D$ are the vev's of the 
scalar field and its dual, surviving the Higgs phenomenon in the spontaneously 
broken phase SU(2) $\ra$ U(1). Classically $a_D = \tau a$, where 
$\tau=\tau_R + i \tau_I = \frac{\theta}{2\pi}+i\frac{4\pi}{g^2}$ is the 
complex coupling, $g$ is the renormalizable SU(2) coupling and $\theta$ is the
CP violating vacuum angle. In the quantum theory $a_D$ is identified with 
$\partial {\cal F} / \partial a$, where $\cal F$ is the holomorphic 
prepotential.

\noindent
In a nutshell the important features of $Z$ in the SW model are:
\begin{enumerate}
\item{It allows for spontaneous symmetry breaking (SSB) of the gauge symmetry 
within the Susy theory;}
\item{It gives the complete and exact mass spectrum for the elementary 
particles as well as the topological excitations;}
\item{It exhibits an explicit SL(2, $\Z$) duality symmetry, which 
is not a symmetry of the theory in the Noether sense;}
\item{In the quantum theory it is the most important global piece of 
information at our disposal on the space of the gauge inequivalent vacua
(the {\it moduli} space ${\cal M}_q$).}
\end{enumerate}
The first two results in the list date back to the early days of the mass
formulae. In fact, in their classic paper Witten and Olive \cite{wo}
proved (a simplified version of) Eq.(\ref{mass}) for a semi-classical N=2 Susy
Georgi-Glashow model with gauge group SO(3). In contrast, the results 3 and 4 
are the crucial ingredients to solve the quantum corrected theory.
It is then not surprising that, following the paper of Seiberg and Witten, 
there have been several attempts to compute the mass formula for the quantum 
case \cite{roc}, \cite{van}. 

\noindent
In fact, in \cite{roc} and \cite{van} 
there is no direct proof of this formula but only a 
brief check that the bosonic terms of the SU(2) high energy effective 
Hamiltonian for a magnetic monopole admit a BPS lower bound given by 
$\sqrt2 |{\cal F}^{(1)}(a)|$ (here and in the following by ${\cal F}^{(n)}$
we mean the $n^{\rm th}$ derivative of ${\cal F}$).
A similar type of BPS computation, only slightly more general, has been 
performed in \cite{roc}. There the authors considered again the SU(2) 
high energy effective Hamiltonian $H$ but this time for a dyon, namely also 
the electric field contribution was considered. 
Of course the topological term $H_{\rm top}$ is the lower bound for $H$, the 
inequality $H \ge H_{\rm top}$ being saturated when the configurations of the 
fields satisfy the BPS equations \cite{bps}. Thus the authors in \cite{roc} 
found that $H_{\rm top} = \sqrt2 |n_e a + n_m {\cal F}^{(1)}(a)|$ therefore 
they identified the r.h.s. with the modulus of the central charge $|Z|$.

\noindent
A common feature of all those computations is that they only consider the 
bosonic contributions to $|Z|$ and this is rather unsatisfactory since, due to
Susy, one might expect fermionic terms to play a role. Furthermore the 
computations briefly described above are rather indirect. The complete and 
direct computation has to involve the Noether supercharges constructed from 
the correspondent Lagrangian. Two independent complete computations appear now
in \cite{ior} and \cite{wolf}. The aim of this paper is to show what we have 
learnt about Susy Noether currents by performing those computations.

\noindent
Susy Noether currents present quite serious difficulties due to the following 
reasons. First Susy is a space-time symmetry therefore the standard procedure 
to find Noether currents does not give a unique answer. A term,
often called {\it improvement}, has to be added to the term one would obtain 
for an internal symmetry. 
The additional term is not unique, it can be fixed only by requiring the
charge to produce the Susy transformations one starts with, and  for non 
trivial theories it is by no means easy to compute.
Second the linear realization of Susy involves Lagrange multipliers
called dummy-fields, which of course have no canonical conjugate momenta.
On the other hand, if dummy-fields are eliminated to produce a standard
Lagrangian, then the variations of the fields are no longer linear and 
the Noether currents are no longer bilinear.
A further problem is that the variations of the fields involve 
space-time derivatives and this happens in a {\it fermion-boson asymmetric} 
way (the variations of the fermions involve derivatives of 
the bosons but not conversely). This implies some double-counting 
solved only by a correct choice of the current.

\noindent
We have solved these problems by implementing a canonical formalism 
in the different cases under consideration. Firstly we construct the Noether
currents for the classical limit of the U(1) sector of the theory. 
In this case Susy is linearly realized 
regardless of the dummy fields, no complications arising in the effective case
are present and the fields are non-interacting. When the procedure is clearly
stated in this case we move to the next level, the effective U(1) sector
and we see what is left from the classical case and what is new. Now the 
currents are very different and, for instance, we cannot use the classical
formula to overcome the above mentioned 
fermi-bose asymmetry in the transformations of the fields.
Nevertheless the constraints imposed by Susy are 
strong enough to force the effective centre to have an identical {\it form} as 
the classical one\footnote{Of course this does {\it not} mean that 
quantum corrections are not present, as is expected to be the case for N=4,
but only that having a {\it dictionary} we could replace classical quantities
by their quantum correspondents with no other changes.}, proving the SW
conjecture that $a_D = {\cal F}^{(1)}(a)$. The last step 
is to consider the SU(2) sector. There we find that the canonical 
procedure implemented in the U(1) sector does not need any further 
change and our analysis confirms that U(1) is the only sector that 
contributes to the centre.

\noindent
Naturally future work would be the generalization of our results to any Susy 
theory, possibly to obtain a general {\it Susy-Noether Theorem}.
The task is by no means easy due to the above mentioned problems
 and other difficulties. For instance, as well known, for 
ordinary space-time symmetries the energy momentum tensor $T_{\mu \nu}$ 
can be obtained by embedding the theory in a curved space-time with metric 
$g_{\mu \nu}$, defining $T_{\mu \nu} = \delta S / \delta g^{\mu \nu}$
and then taking the flat-space limit. In Susy the situation is much more 
complicated because the embedding has to be in a curved {\it super}space,
thus pointing to supergravity which is much more involved than simple 
gravity. 

\noindent
One may also want to investigate the (non-holomorphic) 
next-to-leading order term in the superfield expansion of the SW effective 
Action \cite{weir}. The presence of derivatives higher than second spoils the 
canonical approach and Noether procedure cannot be trivially applied. The 
interest here is to understand how the lack of canonicity and holomorphy (a 
crucial ingredient for the solution of SW model) affects the currents and 
charges, and therefore the whole theory itself. Of course this analysis is 
somehow more general and it could help to understand how to handle the 
symmetries of full effective Actions.

\section{Noether Currents for Space-Time Symmetries}

\noindent
It is a well known fact that the Noether theorem does not give a unique 
conserved current when applied to space-time symmetries. This can be seen 
immediately by defining as {\it symmetry} of a field theory the transformation 
of the fields and/or of the coordinates leaving invariant the Action 
${\cal A} = \int d^4 x {\cal L} (\Phi_i , \partial \Phi_i)$, where $\Phi_i$ 
are fields of arbitrary spin and the invariance is obtained algebraically, 
i.e. without the use of the Euler-Lagrange equations for the fields 
(off-shell). Thus, this definition clearly distinguishes the two cases of 
(rigid) internal symmetry, $\delta {\cal L} = 0$, and space-time symmetry, 
$\delta {\cal L} = \partial^\mu V_\mu$. In the first case no ambiguity arises
and the on-shell conserved current is $N_\mu = \Pi^i_\mu \delta \Phi_i$, 
where $\Pi^i_\mu = \partial {\cal L} / \partial (\partial^\mu \Phi_i)$.
In the second case the current is 
\be\label{n-v}
J_\mu = N_\mu - V_\mu
\ee 
and is clearly not unique since any other $V'_\mu = V_\mu 
+ \partial^\nu W_{[\mu \nu]}$ still satisfies the requirement above introduced.
Often the terms $\partial^\nu W_{[\mu \nu]}$ have to be added {\it ad hoc}, by 
requiring the charges to reproduce the symmetry transformations or, for 
instance, to fit the current into a Susy multiplet. 

\noindent
It is interesting to remember here that there are two Noether theorems 
\cite{emmy}, the first for {\it rigid} symmetry, and the second for {\it local}
symmetry. The theorem we used in the previous discussion is the first theorem,
where the parameter of the infinitesimal transformation $\eps$ is taken to be 
constant. On the other hand one can also produce a Noether current by letting
$\eps$ become local and varying the Action with respect to this parameter 
one obtains
\be\label{jloc}
\delta_{\rm local} {\cal A} = - \int d^4 x J^\mu \partial_\mu \eps 
\ee
where a surface term $\int d^4 x \partial_\mu (J^\mu \eps)$ is discarded. This
alternative method relies anyway on the first Noether theorem since one is 
entitled to identify $J^\mu$ with a conserved current only if the Action is 
invariant under the rigid transformation. Or, alternatively, it is sufficient
to remember that in general even if the Action is locally symmetric the 
(improper, according to Noether's terminology) conservation law descends from 
the rigid symmetry alone. 
Another point is that in general the currents in Eq.(\ref{n-v}) and in 
Eq.(\ref{jloc}) can differ by a $\partial_\nu W^{[\mu \nu]}$ type of term. 
Thus only when internal symmetries are concerned  ($V_\mu = 0$) the two 
alternative procedures give the same answer. On the contrary for space-time
symmetries this is not the case. 
Another problem is the canonical structure of the current\footnote{By 
canonical here we mean that we can write the current as $\Pi^\mu \delta \Phi$,
therefore the charge is $Q=\int d^3 x \Pi \delta \Phi$, where $\Pi = \Pi^0$.
Thus we can immediately generate the transformation of the field $\Phi$ by 
acting on it with its canonically conjugate momentum $\Pi$.}. For space-time 
currents the problem is to express $V_\mu$ canonically and this is exactly 
where the issue of the {\it improvement} terms comes into play (see also 
\cite{wei}).

\noindent
In this paper we want to address the question of Noether currents for a highly
non trivial space-time symmetry: Susy. 
The difficulties of Susy-Noether currents are
\begin{itemize}
\item{Susy is a {\it super}space-time symmetry. Thus it shares with standard 
space-time symmetries all the problems above outlined and, furthermore, 
the non commutative structure of the superspace makes highly non trivial to 
obtain the currents by embedding the theory in a curved superspace. }
\item{Linear realization of Susy involves dummy-fields. This poses some 
delicate questions as when to eliminate those Lagrange multipliers and how to 
treat them in a canonical setting.}
\item{Susy variations involve space-time derivatives in a way not 
symmetrical with respect to fields of different spin. This means that Susy 
maps bosonic fields into fermionic fields and fermionic fields into conjugate
momenta of the bosonic fields. This problem causes some double counting 
essentially solved by partially integrating in the fermionic sector.}
\end{itemize}
For the SW model, the situation is even more complicated due to the following 
problem that closes the list of difficulties encountered in the computations 
illustrated later: 
\begin{itemize}
\item{Effective Lagrangians, even non-Susy.
As we shall see, in SW theory we have to deal with effective Lagrangians and 
renormalization does not constraint the fermionic terms to be bilinear and 
the coefficients of the kinetic terms to be constant and in general this 
is not true.} 
\end{itemize}

\noindent
As a matter of fact, the SW effective Lagrangian is quartic in 
the fermionic fields and has coefficients of the kinetic terms that are 
non-polynomial functions of the scalar field. Because of this, the Noether 
procedure requires a great deal of care. For example we shall encounter equal 
time commutations (Poisson brackets\footnote{See Appendix \ref{poisson}.}) 
between fermions and bosons such as
\be
\{ \psi \; , \; \pi_\phi \} = f(\phi) \psi
\quad{\rm from}\quad \{ \pi_{\bpsi} \; , \; \pi_\phi \} = 0 
\ee
where $f(\phi)$ is a non-polynomial function of the scalar field related 
to the coefficient of the kinetic terms. This reflects the difficulty of 
treating Noether currents in a quantum context \cite{lop}, \cite{hurth}.
Let us note here that in SW theory most of the information on the quantum 
corrections is contained in the expressions of the dummy fields on-shell. Thus
by keeping them explicitly in our Lagrangian we can implement the Noether 
procedure in a similar fashion in both cases, classical and quantum. The
difference will show up when we have to explicitly write down the currents
that generate the transformations, namely when we need to express our charges 
in terms of canonical variables.

\noindent
These problems are addressed in our analysis and we can give here a ``working 
recipe'' we have found:
\begin{itemize}
\item{The Susy-Noether  charge that correctly reproduces the Susy 
transformations is the one obtained from $J^\mu = N^\mu - V^\mu$ where
$\delta {\cal L} = \partial_\mu V^\mu$ and $V^\mu$ has to be extracted 
as it is, i.e. no terms like $\partial_\nu W^{[\nu \mu]}$ have to be 
added. Furthermore $V^\mu$ has to be expressed in terms of momenta and Susy
variations of the fields.}
\item{The variation $\delta {\cal L}$ has to be performed off-shell by the
definition of symmetry. Nevertheless the dummy fields, and {\it only} them,
automatically are projected  on-shell.}
\item{The full current $J^\mu$ contains terms of the form 
$\pi_\psi \delta\psi$, that generate the fermionic transformations. The 
{\it same} term can be written as $\pi_\phi \delta \phi \; + \cdots$ therefore 
it also generates the bosonic transformations. The situation is more 
complicated for effective theories.}
\item{When the effective theory allows for a canonical 
description\footnote{For instance in SW this happens only at the first order
in the momentum expansion of the effective Action.}, the canonical 
commutation relations are preserved even if some of the usual assumptions, 
such as that at equal time all fermions and bosons commute, are incorrect. 
Noether currents at the effective level do not exhibit the same
simple expressions as at the classical level.}
\end{itemize}
\noindent
Of course a recipe is not a final solution and lot of work has to be done to 
fully understand the issue of Susy-Noether currents or more generally 
space-time Noether  currents. Nevertheless our work surely is a guideline in 
this direction and successfully solves the problem of the SW Susy currents 
that we intended to study.

\section{Some Simple Examples}

\noindent
We now want to make our recipe more explicit by applying it to two simple 
cases: the Wess-Zumino (WZ) massive model and the classical N=2 Susy 
Yang-Mills (SYM) model. In the first case we shall show the recipe at work 
for the simplest Susy model with dummy fields coupled to physical fields. 
In the second case we shall re-obtain the classical mass formula of Witten and 
Olive. This formula will be useful for exhibiting the formal resemblance of 
the classical and quantum expressions of the central charge $Z$.

\noindent
In both cases we shall start with a doubled fermionic phase-space and impose 
canonicity by partial integration and by means of Poisson brackets \cite{ior}.
Of course similar results would be obtained by making use of second class 
constraints and the correspondent Dirac brackets \cite{wolf}, 
\cite{dir}, \cite{tyut}.

\subsection{WZ Massive Model}

\noindent
The Lagrangian density and Susy transformations of the fields
for this model \cite{wb} are given by 
\be\label{wz}
{\cal L} = -\frac{i}{2} \psi \not\!\partial {\bar\psi}
-\frac{i}{2} {\bar\psi} \not\!{\bar \partial} \psi
- \partial_\mu A \partial^\mu A^\dagger + F F^\dagger
+ m A F + m A^\dagger F^\dagger - \frac{m}{2} \psi^2 
-\frac{m}{2} {\bar \psi}^2
\ee
and
\bea
\delta A = \sqrt2 \eps\psi & \delta A^\dagger = 
\sqrt2 \beps \bpsi \\
\delta \psi_\alpha  = i \sqrt2 (\sigma^\mu \beps)_\alpha \partial_\mu A 
+ \sqrt2 \eps_\alpha F & \delta \bpsi^{\dalpha} = 
i \sqrt2 (\bsigma^\mu \eps)^{\dalpha} \partial_\mu A^\dag 
+ \sqrt2 \beps^{\dalpha} F^\dag \\
\delta F = i \sqrt2 {\beps}\not\!{\bar \partial} \psi &
\delta F^\dag = i \sqrt2 \eps \not\!\partial {\bpsi}
\eea
where $A$ is a complex scalar field, $\psi$ is its Susy fermionic partner in 
Weyl notation\footnote{See Appendix \ref{not}} and $F$ is the complex bosonic 
dummy field.

\noindent
The phase-space 
$(A,\pi_A ; A^\dag,\pi_A^\dag | \psi,\pi_\psi; \bpsi,\pi_{\bpsi})$,
is doubled fermionic because $\psi$ and $\bpsi$ play the role of fields and 
momenta at the same time. A proper phase-space would use only $(\psi,\pi_\psi)$
or $(\bpsi, \pi_{\bpsi})$. It is really a matter of taste {\it when} to 
integrate by parts to obtain a proper phase-space and implement the canonical 
Poisson brackets. In fact, even if $N^\mu$ and $V^\mu$ both change under
partial integration, the total current $J^\mu$ is {\it formally} invariant,
when expressed in terms of the fields and their derivatives, but not in terms 
of the fields and their momenta.

\noindent
Let us keep (\ref{wz}) as it stands, define the following non canonical momenta
\bea
\pi^\mu_{\psi \alpha} = \frac{i}{2} (\sigma^\mu \bpsi)_\alpha &,&
\pi^{\mu \dalpha}_{\bpsi} = \frac{i}{2} (\bsigma^\mu \psi)^{\dalpha} 
\label{momwz}\\
\pi^\mu_A = -\partial^\mu A^\dagger &,& \pi^\mu_{A^\dagger} = -\partial^\mu A
\eea
and use Eq. (\ref{n-v}) to obtain the supersymmetric current $J^\mu$. 

\noindent
We compute $V^\mu$ by varying (\ref{wz}) off-shell, under the given 
transformations, and obtain\footnote{Here and in the following 
we choose to keep the Susy parameters $\eps$'s, since this simplifies some of 
the computations involving spinors.}
\bea
V^\mu &=& \delta A \pi^\mu_A + \delta A^\dagger \pi^\mu_{A^\dagger} 
        - \delta^A \psi \pi^\mu_\psi 
        - \delta^{A^\dagger} \bar\psi \pi^\mu_{\bar\psi} 
        + \delta^F \psi \pi^\mu_\psi
        + \delta^{F^\dagger} \bar\psi \pi^\mu_{\bar\psi} \nonumber \\
      && -2 \delta^{F_{\rm on}}\psi \pi^\mu_\psi
         -2 \delta^{F^\dag_{\rm on}}\bpsi \pi^\mu_{\bpsi}
\eea
where $\delta^X Y$ stands for the part of the variation of $Y$ which 
contains $X$ (for instance $\delta^F \psi$ stands for $\sqrt2 \epsilon F$) 
and $F_{\rm on}$, $F^\dag_{\rm on}$ are the dummy fields 
given by their expressions on-shell ($F_{\rm on} = - m A^\dag$). 
Note here that we succeeded in finding an expression for $V^\mu$ in terms of 
$\pi^\mu$'s and variations of the fields. Note also that the terms involving 
$F_{\rm on}$ and  $F^\dag_{\rm on}$ were obtained automatically.

\noindent
Then we write the rigid part of the current
\be
N^\mu = 
\delta A \pi^\mu_A + \delta A^\dagger \pi^\mu_{A^\dagger}
+ \delta \psi \pi^\mu_\psi + \delta \bar\psi \pi^\mu_{\bar\psi}
\ee
and the current is given by 
\be\label{fullwz}
J^\mu = N^\mu - V^\mu 
      = 2 {\big (}\delta^{\rm on} \psi \pi_{\psi}^\mu +
	  \delta^{\rm on} \bar\psi \pi_{\bar\psi}^\mu {\big )}
\ee
therefore $J^\mu = 2 (N^\mu)_{\rm fermi}^{\rm on}$, with obvious 
notation. Note that only the variations with $F$ and $F^\dag$ on-shell occur.

\noindent
Therefore we conclude that:  
1) the dummy fields are on-shell automatically and,
if we keep the fermionic non canonical momenta given in (\ref{momwz}), 
2) the full current is given by {\it twice} the fermionic rigid current 
$(N^\mu)_{\rm fermi}^{\rm on}$.

\noindent 
The first (on-shell) result illustrates the second ingredient of the recipe 
given above. We shall see in the highly non trivial case of the SW effective 
Action that the on-shell result still holds, and it seems to be a general 
feature of Susy-Noether currents. 

\noindent 
The second result, instead, is only valid for simple Lagrangians and 
it breaks down for less trivial cases\footnote{There are two reasons for that 
curious result: the fictitious double counting of the fermionic 
degrees of freedom and the the fermi-bose asymmetry in the canonical structure
of the Susy transformations.}.
Nevertheless, when applicable, Eq.(\ref{fullwz}) remains a labour saving 
formula. All we have to do is to rewrite $J^\mu$ in terms of fields 
and their derivatives
\be
J^\mu = \sqrt2 (\bar\psi \bar\sigma^\mu \sigma^\nu \bar\epsilon \partial_\nu A
+ i \epsilon \sigma^\mu \bar\psi F_{\rm on} + {\rm h.c.})
\ee
then choose one partial integration 
\bea
J^\mu &=& \delta_{\rm on}\psi {\pi^\mu}^I_\psi +
\sqrt2 \psi \sigma^\mu \bar\sigma^\nu \epsilon \partial_\nu A^\dagger
+ i \sqrt2 \bar\epsilon \bar\sigma^\mu \psi F^\dagger_{\rm on} \\
{\rm or} &=& 
\sqrt2 \bar\psi \bar\sigma^\mu \sigma^\nu \bar\epsilon \partial_\nu A
+ i \sqrt2 \epsilon \sigma^\mu \bar\psi F_{\rm on} 
+ \delta_{\rm on}\bar\psi {\pi^\mu}^{II}_{\bar\psi}
\eea
where ${\pi^\mu}^I_\psi = i \sigma^\mu \bar\psi$ 
(${\pi^\mu}^{I}_{\bar\psi} = 0$) and 
${\pi^\mu}^{II}_{\bar\psi} = i {\bar\sigma}^\mu \psi$ 
(${\pi^\mu}^{II}_\psi = 0$) are the canonical
momenta obtained by (\ref{wz}) conveniently integrated by parts, 
and perform our computations using canonical Poisson brackets. To integrate
by parts in the effective SW theory a greater deal of care is needed due to 
the fact that the coefficients of the kinetic terms are functions of the 
scalar field.

\noindent
Choosing, for instance, the phase-space 
$(A,\pi_A ; A^\dag,\pi_A^\dag | \psi,\pi^I_\psi)$, what is left is to check 
that the charge
\be\label{qwz}
{\cal Q} \equiv \int d^3 x J^0 (x) = \int d^3 x {\Big (}
\delta_{\rm on}\psi \pi^I_\psi +
\sqrt2 \psi \sigma^0 \bar\sigma^\nu \eps \partial_\nu A^\dagger
+ i \sqrt2 \beps \bsigma^0 \psi F^\dagger_{\rm on} {\Big )}
\ee
correctly generates the transformations. This is a trivial task in this 
case since the current and the expression of the dummy 
fields on-shell are very simple and the transformations can be read off 
immediately from the charge (\ref{qwz}). We shall see that for more complicated
models this is not the case. It is worthwhile to notice at this point that to 
generate the transformations of the scalar field $A^\dag$ one has to use 
\be
\delta^A \psi {\pi^\mu}^I_\psi = \delta A^\dag \pi_{A^\dag} 
+ \sqrt2 \bpsi \bsigma^0 \sigma^i \beps \partial_i A 
\ee
Notice also that the transformation of $\bpsi$ is obtained by acting with the 
charge on the conjugate momentum of $\psi$: 
$\{ {\cal Q} \; , \; \pi^I_\psi \}_{-}$.

\subsection{N=2 SYM Model}

\noindent
There exist two massless N=2 Susy multiplets with maximal helicity 1 or 
less: the vector multiplet and the scalar multiplet \cite{west}, \cite{1001}. 
We are interested in the vector multiplet $\Psi$, also referred to as the N=2 
SYM multiplet, for the moment in its Abelian formulation. Its spin 
content is $(1,\frac{1}{2},\frac{1}{2},0,0)$ and, in terms of physical fields,
it can accommodate 1 vector field $v_\mu$, 2 Weyl fermions $\psi$ and 
$\lambda$, one complex scalar $A$.
The N=2 vector multiplet can be arranged into two N=1 multiplets, the vector 
(or YM) multiplet $W = (\lambda_\alpha, \; v_\mu,  \; D)$ and the scalar 
multiplet $\Phi = (A, \; \psi_\alpha, \; E)$, where $E$ and $D$ are the 
(bosonic) dummy fields\footnote{We use the same symbol $E$ for the electric 
field and for the dummy field. Its meaning will be clear from the context.}, 
related by $R$-symmetry: $\psi \lra -\lambda$, $E^\dag \lra E$ and 
$v^\mu \to -v^\mu$ (charge conjugation).

\noindent
The N=2 Susy transformations of these fields are well known. In our notation 
the first set of transformations is given by \cite{dorey}
\bea
\delta_1 A &=& {\sqrt 2} \eps_1 \psi \nonumber \\
\delta_1 \psi^\alpha &=& {\sqrt 2}\eps_1^\alpha E \label{trns1}\\
\delta_1 E &=& 0 \nonumber 
\eea
\bea
\delta_1 E^\dag &=& i {\sqrt 2} \eps_1 \not\!\partial \bpsi \nonumber \\
\delta_1 \bpsi_{\dalpha} &=& - i {\sqrt2} \eps_1^\alpha 
               \not\!\partial_{\alpha \dalpha} A^\dag \\
\delta_1 A^\dag &=& 0 \nonumber 
\eea
\bea
\delta_1 \lambda^\alpha &=& 
- \eps^\beta_1 ( \sigma^{\mu \nu \: \alpha}_{\: \beta} v_{\mu \nu}
    - i \delta^\alpha_\beta D ) \nonumber \\
\delta_1 v^\mu &=& i \eps_1 \sigma^\mu \blambda \quad , \quad
\delta_1 D = - \eps_1 \not\!\partial \blambda \label{trns2} \\
\delta_1 \blambda_{\dalpha} &=& 0 \nonumber 
\eea
where $v_{\mu \nu} = \partial_\mu v_\nu - \partial_\nu v_\mu$
is the Abelian vector field strength. By $R$-symmetry one can obtain the 
second set of transformations by simply replacing $1 \to 2$, 
$\psi \lra -\lambda$, $v^\mu \to -v^\mu$ and $E^\dag \lra E$ in the first set.

\noindent
The N=2 SYM low-energy effective Lagrangian, up to second derivatives 
of the fields and four fermions is given by \cite{monto}
\bea
{\cal L} &=&
\frac{\rm Im}{4\pi} {\Big (} - {\cal F}^{(2)} (A)
   [\partial_\mu A^\dag \partial^\mu A
    + \frac{1}{4} v_{\mu \nu}{\hat v}^{\mu \nu}
    + i \psi  \not\!\partial {\bar \psi} 
    + i \lambda  \not\!\partial {\bar \lambda}
    - (E E^\dag + \frac{1}{2} D^2)] \nonumber \\
&+& {\cal F}^{(3)}(A) 
    [\frac{1}{\sqrt 2} \lambda\sigma^{\mu\nu}\psi v_{\mu\nu}
      -\frac{1}{2}(E^\dag \psi^2 + E \lambda^2)
      + \frac{i}{\sqrt 2} D \psi \lambda ] 
     + {\cal F}^{(4)} (A)[\frac{1}{4} \psi^2\lambda^2] {\Big )} \label{lu1eff}
\eea
where ${\cal F}(A)$ is a holomorphic and analytic\footnote{By analytic, we 
mean that it can have branch cuts, poles etc., but no essential 
singularities.} function of the scalar field;
$v^*_{\mu \nu}= \eps_{\mu \nu \rho \sigma} v^{\rho \sigma}$ is the dual
of $v^{\mu \nu}$, 
$\hat{v}_{\mu \nu} = v_{\mu \nu} + \frac{i}{2} v^*_{\mu \nu}$ is its self-dual
projection and 
$\hat{v}^\dag_{\mu \nu} = v_{\mu \nu} - \frac{i}{2} v^*_{\mu \nu}$ its 
anti-self-dual projection\footnote{If we define the electric and magnetic 
fields as usual, $E^i=v^{0i}$ and $B^i=\frac{1}{2} \eps^{0ijk}v_{jk}$, 
respectively, we have $\hat{v}^{0i} = E^i + i B^i$ and 
$\hat{v}^{\dag 0i} = E^i - i B^i$.}. Susy constraints all the fields to 
be in the same representation of the gauge group as the vector field, 
namely the adjoint representation. In the U(1) case this representation is
trivial, and the derivatives are standard rather than covariant. We notice 
here that $v_0$ plays the role of a Lagrange multiplier, and the associate 
constraint is the Gauss law. Thus, by taking the derivative of $\cal L$ with 
respect to $v_0$ we obtain the quantum modified Gauss law for this theory, 
namely
\be\label{gauss1}
0 = \frac{\partial {\cal L}}{\partial v_0} = \partial_i \Pi^i
\ee
where $\Pi^i = \partial {\cal L} / \partial(\partial_0 v_i)$ is the conjugate
momentum of $v_i$, given by
\be
\Pi^i = -({\cal I} E^i - {\cal R} B^i)
+ \frac{1}{i \sqrt2} ({\cal F}^{(3)} \lambda\sigma^{0i}\psi 
- {{\cal F}^{(3)}}^\dag \blambda \bsigma^{0i} \bpsi) 
\ee
and ${\cal F}^{(2)} = {\cal R} + i {\cal I}$.

\noindent
The theory described by the Lagrangian in (\ref{lu1eff}) is the Abelian 
sector of the SW model. Their achievement essentially consists in the exact 
determination of the function $\cal F$ in the three sectors of the space of 
gauge inequivalent vacua ${\cal M}_q$. 

\pagebreak
\noindent
{\it U(1) Classical Model}

\noindent
We shall study, for the moment, the classical limit of this 
Lagrangian\footnote{At this end it is sufficient to recall that in the 
classical case there is no running of the coupling, therefore there 
is only one global description at any scale of the energy. Thus we can use 
the expression 
\be\label{fu1}
{\cal F} (A) = \frac{1}{2} \tau A^2 
+ A^2 \frac{i \hbar}{2 \pi} \ln (\frac{A^2}{\Lambda^2})
+ A^2 \sum_{k=1}^{\infty} c_k \frac{\Lambda^{4k}}{A^{4k}}
\ee
to write the classical limit as
\be\label{clim}
{\cal F}(A) \to \frac{1}{2} \tau A^2
\ee
where $\tau$ is the complex coupling constant already introduced.}.
In this limit the second line of (\ref{lu1eff}) vanishes, and  writing 
explicitly, the first line becomes
\bea
{\cal L} & = & 
- \frac{1}{g^2}  \Big{(} \frac{1}{4} v_{\mu \nu} v^{\mu \nu}
+ \partial_\mu A^\dagger \partial^\mu A
- (E E^\dag + \frac{1}{2} D^2) \Big{)}
- \frac{\theta}{64 \pi^2}v_{\mu \nu} v^{* \mu \nu} \nonumber \\
&& -\frac{1}{4\pi}  \Big{(}
\frac{\tau}{2} \psi  \not\!\partial {\bpsi}
- \frac{\tau^*}{2} {\bpsi} {\not\!\bar\partial} \psi
+ \frac{\tau}{2} \lambda  \not\!\partial {\blambda}
- \frac{\tau^*}{2} {\blambda} {\not\!\bar\partial} \lambda \Big{)} 
\label{lu1class}
\eea
If we keep the improper phase-space 
$(A,\pi_A ; A^\dag,\pi_{A^\dag} ; v_i, \Pi^i| 
\psi,\pi_\psi; \bpsi,\pi_{\bpsi} ; \lambda,\pi_\lambda ; 
\blambda,\pi_{\blambda})$, as in the WZ model, 
the non canonical momenta are given by
\be
\pi^\mu_A = - \frac{1}{g^2} \partial^\mu A^\dagger = (\pi^\mu_{A^\dag})^\dag 
\; , \;
\Pi^{\mu \nu}  
= -\frac{1}{8 \pi i} (\tau \hat{v}^{\mu \nu} - \tau^* \hat{v}^{\dag \mu \nu})
\ee
and
\be
4 \pi (\pi^\mu_{\bar \psi})_{\dot\alpha} =
 \frac{\tau}{2} \psi^\alpha \sigma^\mu_{\alpha \dot\alpha} \quad , \quad
4 \pi (\pi^\mu_{\psi})^{\alpha} =
 -\frac{\tau^*}{2} \bar\psi_{\dot\alpha} \bar\sigma^{\mu \dot\alpha \alpha}
\ee
similarly for $\lambda$.

\noindent
As explained earlier in order to compute the Susy Noether 
currents\footnote{We have only to compute the first Susy current, since by 
R-symmetry, charge conjugation and complex conjugation we can obtain the 
other currents.} we have to compute their $V_\mu$ part. In the classical case 
this is an easy matter, but for the effective theory it is not trivial. 
We obtain 
\be \label{V1class}
V_1^\mu = \pi_A^\mu \delta_1A + \Pi^{\mu \nu}\delta_1 v_\nu 
+ \frac{\tau^*}{\tau} \delta_1 {\bar \psi} \pi_{\bar\psi}^\mu
+ \delta_1 \psi \pi_\psi^\mu
+ \delta_1^D \lambda \pi_{\lambda}^\mu
+ \frac{\tau }{\tau^*} \delta_1^v \lambda \pi_{\lambda}^\mu 
\ee
where $\delta_1 {\cal L} = \partial_\mu V_1^\mu$, and again $\delta^X Y$ 
stands for the term in the variation of $Y$ that contains $X$ (for instance 
$\delta_1^D \lambda^\alpha \equiv i \eps_1^\alpha D$).
The total current $J_1^\mu$ is then given by\footnote{$J_1^\mu$ stands for 
$\eps^1 J_1^\mu$ {\it or} $\eps_1 J^{1 \mu}$. In the following we shall
not keep track of the position of these indices, they will be simply treated 
as labels.}
\bea
J_1^\mu &=& N^\mu_1 - V_1^\mu \nonumber \\
        &=& \pi_A^\mu \delta_1A + \Pi^{\mu \nu}\delta_1 v_\nu 
          + \delta_1 \psi \pi_{\psi}^\mu 
          + \delta_1 \lambda \pi_{\lambda}^\mu
          + \delta_1 \bar\psi \pi_{\bar\psi}^\mu - V_1^\mu \nonumber \\
	&=& \frac{2i\tau_I}{\tau} \delta_1 \bar\psi \pi_{\bar\psi}^\mu
 - \frac{2i\tau_I}{\tau^*} \delta_1^{\rm on} \lambda \pi_{\lambda}^\mu 
\eea
where $N^\mu_1$ is the rigid current, $\delta_1 {\blambda} = 0$ and 
$\delta_1^{\rm on} \lambda$ stand for the variation of $\lambda$ with dummy 
fields on-shell\footnote{In this case this means $E=D=0$ and one could also 
wonder if they are simply cancelled in the total current. But, in agreement 
with our recipe, we shall see later that indeed the dummy fields, and 
{\it only} them, have been automatically projected on-shell.} (there are no 
dummy fields in the variation of $\bpsi$).

\noindent
If we set $\theta = 0$ in this non-canonical setting, we recover the same type
of expression, $J^\mu = 2 (N^\mu_{\rm fermi})^{\rm on}$, for the total 
current obtained in the massive WZ model, namely
\be
J_1^\mu|_{\theta = 0} =  2 (\delta_1^{\rm on} \lambda \pi_{\lambda}^\mu 
	  + \delta_1 \bpsi \pi_{\bpsi}^\mu)
\ee
We see again that the double counting of the fermionic degrees of freedom 
provides a very compact formula for the currents. All the information is 
contained in the fermionic sector, since the variations of the fermions 
contain the bosonic momenta. Unfortunately this does not seem to be the case
for the effective theory, as we shall see in the next Section.

\noindent
We have now to integrate by parts in the fermionic sector\footnote{In this 
case there is no effect of the partial integration on the bosonic momenta 
since $\partial_\mu \tau = 0$, we shall see that this is no longer the case 
for the effective theory.} of the Lagrangian (\ref{lu1class}) to obtain 
a proper phase space. 
Everything proceeds along the same lines as for the WZ model. 
Thus the canonical fermionic momenta are 
\be
(\pi^{I \mu}_{\bpsi})_{\dalpha} =
  \frac{i}{g^2} \psi^\alpha \sigma^\mu_{\alpha \dalpha} \;,\;
(\pi^{I \mu}_{\blambda})_{\dalpha} =
  \frac{i}{g^2} \lambda^\alpha \sigma^\mu_{\alpha \dalpha}
\ee
and $\pi^{I \mu}_\psi = \pi^{I \mu}_\lambda =0$, where with $I$ we indicate 
one of the two possible choices ($\bpsi$ and $\blambda$ are the fields).
The partial integration changes $V^\mu$, but, of course, also $N^\mu$ changes 
accordingly and they still combine to give the same total current $J^\mu$. 
Namely 
$
N^{I \mu}_1 =  \pi^\mu_A \delta_1 A  + 
\Pi^{\mu \nu} \delta_1 v_\nu + \delta_1 \bpsi \pi^\mu_{\bpsi} 
$
and 
$
V_1^{I \mu} =   \pi^\mu_A \delta_1 A
+ \frac{1}{8 \pi i} \tau^* \eps_1 \sigma_\nu \blambda v^{* \mu \nu} 
$
that give a total current 
\be\label{jclasstot}
J_1^{I \mu}
= \delta_1 \bpsi \pi_{\bpsi}^{I \mu} 
-\frac{i}{g^2}\eps_1 \sigma_\nu \blambda {\hat v}^{\mu \nu} 
= -\sqrt2 \eps_1 \sigma^\nu \bsigma^\mu \psi\pi_{\nu A}
+ \Pi^{\mu \nu} \delta_1 v_\nu
- \frac{1}{8 \pi i} \tau^* \eps_1 \sigma_\nu \blambda v^{* \mu \nu}
\ee
where we used the identities
\bea
\delta_1 \bpsi \pi_{\bpsi}^{I \mu} &=& 
-\sqrt2 \eps_1 \sigma^\nu \bsigma^\mu \psi \pi_{\nu A} \label{id1}\\
-\frac{i}{g^2}\eps_1 \sigma_\nu \blambda {\hat v}^{\mu \nu} &=&
\Pi^{\mu \nu} \delta_1 v_\nu
- \frac{1}{8 \pi i} \tau^* \eps_1 \sigma_\nu \blambda v^{* \mu \nu} \label{id2}
\eea
The point we make here is that the current is once and for all given by
\be
J_1^\mu = \frac{1}{g^2}
(\sqrt2 \eps_1 \sigma^\nu \bsigma^\mu \psi \partial_\nu A^\dag 
-i \eps_1 \sigma_\nu \blambda {\hat v}^{\mu \nu})
\ee
but its content in terms of canonical variables changes according to partial 
integration. Furthermore one has to conveniently re-express the current 
obtained via Noether procedure to obtain the relevant expression in terms of 
bosonic or fermionic momenta and transformations. Note also that $\theta$ does
not appear in the explicit formula, as could be expected.

\noindent
Choosing the temporal gauge for the vector field, $v^0 = 0$, and defining the 
canonical momenta as usual (remember that our metric is  
$\eta^{\mu \nu} = {\rm diag}(-1, 1, 1, 1)$) we can write down the first Susy 
charge $Q_{1 \alpha}$ and the other charges are simply obtained by R-symmetry,
charge conjugation and complex conjugation. This charges correctly reproduce
the Susy transformations.

\noindent
We can now compute the Poisson brackets that contribute to the centre
$
Z = \frac{i}{4} \eps^{\alpha \beta} \{ Q_{1 \alpha} , Q_{2 \beta} \}_{+}
$
obtaining 
\be
Z = \int d^3 x {\Big (} \partial_i [i \sqrt2 
(\Pi^i A^\dag + \frac{1}{4\pi} B^i A^\dag_D) 
- \frac{1}{4\pi} \tau^* \bpsi \bsigma^{i 0} \blambda ]
+ i \sqrt2 [(\partial_i \Pi^i) A^\dag + 
\frac{1}{4\pi} (\partial_i v^{* 0i}) A^\dag_D] {\Big )} 
\ee
where $A_D = \tau A$.
By using the Bianchi identities, $\partial_i v^{* 0i} = 0$, and the classical 
limit of the Gauss law (\ref{gauss1}), we are left with a total divergence. 
The final expression for $Z$ is then given by
\be\label{zclass}
Z= i \sqrt2 \int d^2 \vec{\Sigma} \cdot 
(\vec\Pi A^\dag + \frac{1}{4\pi} \vec{B} A^\dag_D)
\ee
where $d^2\vec\Sigma$ is the measure on the sphere at infinity $S^2_\infty$, 
and we have made the usual assumption that $\bpsi$ and $\blambda$ fall off 
at least like $r^{-{3\over 2}}$. This is the U(1) version of the well known 
result of Witten and Olive. Note that we ended up with the anti-holomorphic 
centre.

\noindent
{\it SU(2) Classical Model}

\noindent
Before moving on to the quantum theory we want to make full contact with 
Witten and Olive's computation that was performed for the classical non 
abelian model. At this end we see from the current in (\ref{jclasstot}) that a
straightforward generalization to the gauge group SU(2) leads to the 
following charge
\be
\eps_1 Q_1^{\rm SU(2)}
= \int d^3 x (\delta_1 \bpsi^a \pi_{\bpsi}^{a} 
-\frac{i}{g^2}\eps_1 \sigma_i \blambda^a {\hat v}^{a 0 i} )
\ee
where the $a$ is the SU(2) index, we dropped the index $I$, the identities 
(\ref{id1}) and (\ref{id2}) have to be used when necessary and the SU(2) 
momenta and fields are defined as usual (see Appendix \ref{su2conv}). 
By using the same techniques as in the U(1) case, we see that this charge 
correctly generates the transformations that do not involve dummy fields,
namely
$
\delta_1 v^a_i \; , \; 
\delta_1 A^a \quad \delta_1 A^{\dag a} \; , \;
\delta_1 \bar\psi^a \; , \;{\rm and} \;\;
\delta_1 \bar\lambda^a \; , \;
$
but does not generate the transformations involving the dummy fields 
(see, for instance \cite{dorey} and Appendix \ref{su2conv}). Thus some terms 
must be missing. We can obtain the missing terms by considering the classical 
(microscopic) SU(2) Lagrangian ${\cal L}_{\rm class}^{\rm SU(2)}$ and solving 
the Euler-Lagrange equations for $D^a$. The result is given by 
\be
(D^a)^{\rm on}_{\rm class} = i \eps^{a b c} A^b A^{c \dag}
\ee
where we used the standard expression for ${\cal L}_{\rm class}^{\rm SU(2)}$
(see, for instance, \cite{bil} and \cite{dorey}). Note that 
$(E^a)_{\rm class}^{\rm on} = 0$.
From our recipe, we know that the charge has to produce the transformations 
with the dummy fields on shell. $D^a$ appears in the transformation of 
$\lambda^a$ therefore we want to produce $\delta^{D}_1 \lambda^a$ 
from $\{ \epsilon_1 Q^{\rm SU(2)}_1, \pi^a_{\blambda} \}$, where 
$\pi^a_{\blambda} = i \tau_I \lambda^a  \sigma^0$ is the classical SU(2) 
conjugate momentum of $\blambda^a$. Thus we conclude that a missing term in 
the classical charge is given by
\be\label{dclassical}
i  \tau_I \eps_1 \sigma^0 \blambda^a 
\eps^{a c d} A^c A^{d \dag}
\ee
Furthermore this term is the only missing term, because once it is added then 
we obtain all the correct Susy transformations. Thus the final expression
for the classical SU(2) first Susy charge is 
\be
\eps_1 Q_1^{\rm SU(2)}
= \int d^3 x (\delta_1 \bpsi^a \pi_{\bpsi}^{a} 
-\frac{i}{g^2}\eps_1 \sigma_i \blambda^a {\hat v}^{a 0 i} 
+ i  \tau_I \eps_1 \sigma^0 \blambda^a 
\eps^{a c d} A^c A^{d \dag})
\ee
If we compute $Z$ by taking the Poisson brackets of this charge with its 
R-symmetric counterpart we obtain an expression in any respect 
similar\footnote{Of course the Gauss law has to be modified. See more on this
in the next Section.} to (\ref{zclass}) but this time with all the SU(2) 
contributions. By breaking the gauge symmetry along one direction, say 
$<0|A^a|0> = \delta^{a 3} a$, we define the electric and magnetic charges 
{\it \`a la} Witten and Olive \cite{wo},
\bea
n_e &\equiv& \frac{1}{a} \int d^2 \vec{\Sigma} \cdot \vec{\Pi}^3 A^3 \\ 
n_m &\equiv& \frac{1}{a} \int d^2 \vec{\Sigma} \cdot \frac{1}{4\pi} 
\vec{B}^3 A^3 
= \frac{1}{a_D} \int d^2 \vec{\Sigma} \cdot \frac{1}{4\pi} 
\vec{B}^3 A^3_D
\eea
where $a_D = \tau a$ and only the U(1) fields remaining massless after SSB 
appear.

\noindent
We finally have 
\be
Z = i \sqrt2 (n_e a^* + n_m a^*_D)
\ee

\section{Quantum Case: Seiberg-Witten Mass Formula}

\noindent
After these preliminaries, we now wish to consider the quantum SW model.
We shall first concentrate on the U(1) sector and then generalize our results
to the full high energy SU(2) sector.

\subsection{The U(1) Sector}

\noindent
We have now to consider the quantum corrected U(1) Lagrangian given in 
(\ref{lu1eff}).
This time the dummy fields couple non trivially to the fermions. Their 
expressions on-shell are given by 
\be \label{dummy}
D=-\frac{1}{2 \sqrt 2} (f \psi\lambda + f^{\dag} \bpsi\blambda) 
\quad {\rm and} \quad 
E= \frac{i}{4} (f^{\dag} \blambda^2 - f \psi^2) 
\ee
where $f(A,A^\dag) \equiv {\cal F}^{(3)} / {\cal I}$. As we shall see in a 
moment, they represent the most important difference between the classical and
the quantum case.

\noindent
As before, we can concentrate on the computation of the first Susy current 
$J_1^\mu$. The task, of course, is to find $V_1^\mu$ and it turns out that its 
computation in the quantum case is by no means easy. We have found by direct 
computation $V_1^\mu$ and $\bar{V}_1^\mu$. Of course the first one has been 
the most difficult to find, since if one understands how to proceed in the 
first case, the other cases become only tedious checks. We do not want to 
explicitly show here all the details of this lengthy computation. In 
Appendix \ref{compV}, we shall give in some details only the simplest part 
of the computation of $V^\mu_1$, namely the contribution coming from the 
$\cal F$-terms. 

\noindent
If we keep the improper phase-space 
$(A,\pi_A ; A^\dag,\pi_{A^\dag} ; v_i, \Pi^i| 
\psi,\pi_\psi; \bpsi,\pi_{\bpsi} ; \lambda,\pi_\lambda ; 
\blambda,\pi_{\blambda})$, the non canonical momenta are given by
\be\label{ncm1}
\pi^{\mu}_A = - {\cal I} \partial^\mu A^\dag = (\pi^{\mu}_{A^\dag})^\dag
\ee
\be
\Pi^{\mu \nu} = -\frac{1}{2i} ({\cal F}^{(2)} {\hat v}^{\mu \nu} 
- {\cal F}^{(2) \dag} {\hat v}^{\dag \mu \nu}) 
+ \frac{1}{i \sqrt2} ({\cal F}^{(3)} \lambda\sigma^{\mu \nu}\psi 
- {\cal F}^{(3) \dag} \blambda \bsigma^{\mu \nu} \bpsi) 
\ee
and 
\be\label{ncm2}
(\pi^\mu_{\bpsi})_{\dalpha} = \frac{1}{2} {\cal F}^{(2)}
  \psi^\alpha \sigma^\mu_{\alpha \dalpha} \; \; , \; \;
(\pi^\mu_{\psi})^{\alpha} =
  - \frac{1}{2} {\cal F}^{(2) \dag}
  \bpsi_{\dalpha} \bsigma^{\mu \dalpha \alpha} 
\ee
similarly for $\lambda$. Note that for the moment we scale $\cal F$ by a factor
$4 \pi$. The result of the full computation of $V_1^\mu$ ($\cal F$-terms and 
${\cal F}^\dag$-terms) is 
\be
V_1^\mu = \delta_1 A \pi^\mu_A 
+  \frac{{\cal F}^{(2) \dag}}{{\cal F}^{(2)}} \delta_1 \bpsi \pi^\mu_{\bpsi} 
+ \delta_1 \psi \pi^\mu_\psi 
+ \delta_1 \lambda \pi^\mu_\lambda \nonumber
+ \frac{1}{2i} {\cal F}^{(2) \dag} \eps_1 \sigma_\nu \blambda v^{* \mu \nu}
+ \frac{1}{2\sqrt2} {\cal F}^{(3) \dag} \eps_1 \sigma^\mu \bpsi \blambda^2 
\ee
Of course the rigid current is formally identical to the classical one,
$
N^\mu_1=\pi_A^\mu \delta_1A + \Pi^{\mu \nu}\delta_1 v_\nu 
+ \delta_1 \psi \pi_{\psi}^\mu + \delta_1 \lambda \pi_{\lambda}^\mu
+ \delta_1 \bar\psi \pi_{\bar\psi}^\mu
$, 
but the ${\cal F}^{(3) \dag}$ term in $V_1^\mu$ has no classical analogue.
Thus, even if we conveniently rearrange the terms in the total current 
$J_1^\mu = N^\mu_1 - V_1^\mu$ to write
\be
J_1^\mu =  \frac{2 i {\cal I}}{{\cal F}^{(2)}} \delta_1 \bpsi 
           \pi^\mu_{\bpsi}
         -  \frac{2 i {\cal I}}{{\cal F}^{(2) \dag}} \delta_1^{\rm on} 
	    \lambda \pi^\mu_{\lambda}
         +  \frac{1}{2\sqrt2} {{\cal F}^{(3) \dag}} \eps_1 \sigma^\mu 
              \bpsi \blambda^2
         + \frac{1}{\sqrt2} {\cal F}^{(3)} \eps_1 \psi 
           \lambda \sigma^\mu \blambda \label{j12}
\ee
we see that the classical formula, $J_\mu=2(N^\mu_{\rm fermi})^{\rm on}$, no 
longer holds. Therefore we can just move on to a proper phase space, by 
partially integrating in the fermionic sector. Some care is needed due to 
the fact that now the coefficient of the kinetic terms is a function and 
not a constant. By choosing, for instance, $\bpsi$ and $\blambda$ as fields, 
the canonical momentum of $A^\dag$ becomes 
$
\pi^{I \mu}_{A^\dag} = - {\cal I} \partial^\mu A 
- \frac{1}{2} {\cal F}^{(3) \dag} (\bpsi \bsigma^\mu \psi
+ \blambda \bsigma^\mu \lambda) \label{pbI}
$,
and of course the fermionic momenta become
$
(\pi^{I \mu}_{\bpsi})_{\dalpha} = i{\cal I} \psi^\alpha 
\sigma^\mu_{\alpha \dalpha}
$,
$
(\pi^{I \mu}_{\psi})^{\alpha} = 0  \label{pfI}
$,
similarly for $\lambda$. Note that $\Pi^{\mu \nu}$ and $\pi^\mu_A$ do not 
change. 

\noindent
The first Susy charge for the U(1) effective theory\footnote{As in the 
classical case we fix the temporal gauge for the vector field $v^0=0$.} is 
then given by
\be
\eps_1 Q^I_1 = \int d^3 x {\big [} \delta_1 \bar{\psi} 
             \pi^{I}_{\bar\psi} +  \Pi^i \delta_1 v_i
          -  \frac{1}{2i} {{\cal F}^{(2) \dag}} \eps_1 \sigma_i 
              \blambda v^{* 0 i}
          -  \frac{1}{2\sqrt2} {{\cal F}^{(3) \dag}} \eps_1 \sigma^0 
              \bpsi \blambda^2  {\big ]} \label{QI}
\ee
Again we have to conveniently re-express it in terms of fermionic or bosonic 
variables when necessary. 

\noindent
In this case it is not immediate to verify if this charge correctly reproduces 
the Susy transformation. In fact one could wonder if the presence of cubic 
fermionic terms and third derivatives of $\cal F$ would spoil the simple 
structure of the transformations generated by this charge \cite{hurth}. 
This point is really delicate, since the most crucial requirement of the SW
model is that Susy is preserved at quantum level. We leave to Appendix 
\ref{vertrns} the explicit proof that this charge generates exactly the 
transformations given\footnote{In Appendix \ref{vertrns} we shall verify that 
the charge $\eps_1 Q_1^I$, obtained by using $\bpsi$ and $\blambda$ as fields,
works. We do not show there that also the other charge $\eps_1 Q_1^{II}$, 
with $\psi$ and $\lambda$ as fields work, but this is indeed the case. 
This prove that also at the effective level the final result is insensitive 
to partial integration.} in (\ref{trns1})-(\ref{trns2}). The only difference 
between classical and quantum Susy transformations is the expression of the 
dummy fields on-shell: classically they are all zero (in the U(1) sector), at 
the quantum level they are given by Eq. (\ref{dummy}). Note also that in the 
effective case some of the standard assumptions about Poisson brackets do no 
longer hold, as can be seen by the following non trivial bracket
\be
\{ \pi_A \;,\; \pi_{\bpsi} \}_{-} = 0 \; \Rightarrow \;
\{ \pi_A \;,\; \psi \}_{-} = -\frac{i}{2} f \psi 
\ee
Nevertheless the canonical structure survives (see Appendix \ref{poisson}).

\noindent
We are now in the position to compute the central charge for this theory:
\be
Z = \int d^3 x {\Big (} \partial_i [i \sqrt2 \Pi^i A^\dag 
+ \frac{1}{4\pi} B^i {{\cal F}^{(1)}}^\dag
- \frac{1}{4 \pi} {{\cal F}^{(2)}}^\dag
\bpsi \bsigma^{i0} \blambda] 
+ i \sqrt2 [(\partial_i \Pi^i) A^\dag
+ \frac{1}{4\pi} (\partial_i v^{* 0i}) {{\cal F}^{(1)}}^\dag] {\Big )}
\ee
where we reintroduced the factor $4\pi$ and used the formula 
$
Z=\frac{i}{4} \epsilon^{\alpha \beta} \{ Q_{1 \alpha} , Q_{2 \beta} \}_{+}
$.
Imposing the Bianchi identities and the Gauss law (\ref{gauss1}), and 
dropping the fermionic term as in the classical case, we can eventually write
\be\label{zeffu1}
Z = i \sqrt2 \int d^2 \vec{\Sigma} \cdot 
(\vec{\Pi} A^\dag + \frac{1}{4\pi} \vec{B} {\cal F}^{(1) \dag})
\ee
where $B^{i} = \frac{1}{2} \eps^{0ijk} v_{jk}$ as in the classical case. 
Note that from Eq.(\ref{zeffu1}) we can define the SW dual of the scalar 
field $A^\dag$ as given by
\be
A^\dag_D \equiv {{\cal F}^{(1)}}^\dag (A^\dag)
\ee
Surprisingly enough the expression (\ref{zeffu1}) is {\it formally} identical 
to the classical one given in (\ref{zclass}). 
We see that the topological nature of $Z$ is sufficient to protect its form at
the quantum level. All one has to do is to use a little dictionary and replace
classical quantities with their quantum counterparts.

\noindent
Thus we can apply exactly the same logic as in the classical case and define 
the electric and magnetic charges {\it \`a la} Witten and Olive. The final 
expression for the U(1) effective central charge is  
\be
Z = i \sqrt2 (n_e a^* + n_m a^*_D)
\ee
where $<0|A^\dag|0> = a^*$, $<0|{\cal F}^{(1) \dag}|0> = a_D^*$ and $n_e$, 
$n_m$ are the electric and magnetic quantum numbers, respectively.

\noindent
Eventually we proved the SW mass formula. At this end we can simply use the 
BPS type of argument given in \cite{sw} or \cite{roc}, noticing that 
our direct computation includes fermions but they occur as a total divergence
which falls off fast enough to give contribution on $S^2_\infty$. Thus 
\be
M = |Z| = \sqrt2 |n_e a + n_m {\cal F}^{(1)}(a)|
\ee

\noindent
A last remark is now in order. The U(1) low energy theory is invariant under
the linear shift ${\cal F} (A) \to {\cal F} (A) + c A$. This produces an 
ambiguity in the definition of $Z$. For this and other purposes we want also 
to analyse the SU(2) high energy theory.

\subsection{The SU(2) Sector}

\noindent
We wish to generalize the U(1) charge (\ref{QI}) to the SU(2) case. We recall
that for the classical version the transition U(1) $\to$ SU(2) required 
also the addition of the term (\ref{dclassical}). Clearly in the effective 
case this term has to be 
\be
i  {\cal I}^{a b} \epsilon_1 \sigma^0 \bar\lambda^b 
\epsilon^{a c d} A^c A^{d \dagger}
\ee
and it reduces to the classical term (\ref{dclassical}) in the limit
${\cal F}(A) \to \frac{1}{2} \tau A^2$.
The addition of this term suffices also in the quantum theory. In other words
the full SU(2) quantum first\footnote{As usual, the other charges are obtained 
by means of R-symmetry, charge conjugation and complex conjugation.} Susy 
charge is 
\be
\eps_1 Q_1 = 
 \int d^3 x {\Big(} \Pi^{a i} \delta_1 v^a_i
+ \delta_1  {\bpsi}^a \pi^a_{\bpsi} 
+ \frac{i}{2}  {\cal F}^{\dag a b} 
\eps_1 \sigma_i {\blambda}^a v^{* 0 i b} 
- \frac{1}{2\sqrt2}  {\cal F}^{\dag a b c}
\eps_1 \sigma^0 {\bpsi}^a \blambda^b \blambda^c
+ i{\cal I}^{a b} \eps_1 \sigma^0 \blambda^b 
\eps^{a c d} A^c A^{d \dag} {\Big )} \label{Qsu21}
\ee
where we have dropped the label ``SU(2)''. 

\noindent
This term is the only new term required to produce the Susy transformations, 
the Higgs and Yukawa potential in the Hamiltonian and, as we shall see in a 
moment, it is responsible for most of the new terms in the centre.

\noindent
We want now to compute the Hamiltonian $H$ with our charges. The main point 
here is to to obtain the non trivial Gauss law for the SU(2) theory from the
Legendre transformed of $H$. This will be crucial to obtain the correct centre
for this sector of the theory.

\noindent
The formula for $H$ is 
$
H = - \frac{i}{4} \bsigma^{0 \dalpha \alpha}
\{ Q_{1 \alpha} \; , \; \bar{Q}_{1 \dalpha} \}_{+}
$
where we defined $H = P^0 = -P_0$.
This lengthy computation is illustrated in some details in Appendix 
\ref{su2comp}. Its final result is given by
\bea
H &=& \int d^3x {\Big (} -\frac{1}{2}({\cal I}^{ab})^{-1} \Pi^{ai} \Pi^{bi}
- ({\cal I}^{ab})^{-1} {\cal R}^{bc} \Pi^{ai} B^{i c} 
- \frac{1}{2}({\cal I}^{ab})^{-1}{\cal F}^{\dag ab}{\cal F}^{ef}
B^{i b}B^{i f} \nonumber \\
&&-({\cal I}^{ab})^{-1} \pi^a_A (\pi^b_A)^\dag 
+ {\cal I}^{ab} ({\cal D}^i A^a)({\cal D}^i A^{\dag b}) \nonumber \\
&& + i {\cal I}^{ab} \psi^a\sigma^i{\cal D}_i\bpsi^b 
+ i {\cal I}^{ab} \lambda^a\sigma^i{\cal D}_i\blambda^b 
+ \frac{1}{2} (\partial_i {\cal F}^{\dag ab}) \blambda^a \sigma^i \lambda^b 
\nonumber \\
&& - \frac{1}{\sqrt2}{\cal I}^{ad}\eps^{abc}  
(A^c \bpsi^d \blambda^b + A^{\dag c} \psi^d \lambda^b) 
+ \frac{1}{2}{\cal I}^{ab}\eps^{acd} \eps^{bfg}
A^c A^{\dag d} A^f A^{\dag g} \nonumber \\
&& +\frac{i}{\sqrt2} ({\cal I}^{af})^{-1} {\cal F}^{feg} 
\psi^e\sigma_{i0}\lambda^g (\Pi^{ia}+ {\cal F}^{\dag ab} B^{ib}) \nonumber \\
&& -\frac{i}{\sqrt2} ({\cal I}^{ec})^{-1}{\cal F}^{\dag abc} 
\bpsi^a\bsigma_{i0}\blambda^b (\Pi^{ie}+ {\cal F}^{ed} B^{id}) \nonumber \\
&& + \frac{1}{16} {\cal F}^{\dag efg} {\cal F}^{cad} ({\cal I}^{gc})^{-1}
\bpsi^e\bpsi^f \psi^a\psi^c 
+\frac{1}{16} {\cal F}^{\dag abc} {\cal F}^{efg} ({\cal I}^{ae})^{-1}
\blambda^b\blambda^c \lambda^f\lambda^g \nonumber \\
&& + \frac{3}{16} {\cal F}^{\dag bec} {\cal F}^{\dag efg} ({\cal I}^{ab})^{-1}
\bpsi^a\bpsi^c \blambda^f\blambda^g 
+ \frac{3}{16} {\cal F}^{bec} {\cal F}^{efg} ({\cal I}^{ab})^{-1}
\psi^a\psi^c \lambda^f\lambda^g \nonumber \\ 
&& -\frac{1}{2i} 
(\frac{1}{4}{\cal F}^{abcd} \psi^a\psi^b \lambda^c\lambda^d
- \frac{1}{4}{\cal F}^{\dag abcd} \bpsi^a\bpsi^b \blambda^c\blambda^d) 
{\Big)} \nonumber \\
&& + \int d^3x \partial_i 
(\frac{1}{2} {\cal F}^{\dag ab} \blambda^a \bsigma^i \lambda^b
- \frac{i}{2} {\cal I}^{ab} \bpsi^a \bsigma^i \psi^b) \label{Hsu2}
\eea
where $E^{a i} = v^{a 0 i}$ and $B^{a i} = \frac{1}{2} \eps^{0ijk} v^a_{jk}$ 
are the SU(2) generalization of the electric and magnetic fields, respectively.
We notice here that, in the last line, we kept a total divergence to 
explicitly show that we partially integrated the fermionic kinetic terms, in 
order to fix the phase space $(\bpsi, \blambda ; \pi_{\bpsi} , \pi_{\blambda})$
we started with\footnote{It turns out that this total divergence is not 
symmetric with respect to $\psi$ and $\lambda$ and this is reflected in the 
last term of the third line where only $\lambda$-terms appear. This means 
that the Lagrangian we shall obtain by Legendre transforming $H$ will be 
slightly different from the one expected. Nevertheless the difference will 
not affect the conjugate momenta, therefore the Susy charges above 
constructed are not affected by this asymmetry. Furthermore this problem is 
entirely due to the non trivial partial integration in the effective case. As 
explained earlier this does not affect the explicit expression of the currents
and charges.}.

\noindent
By Legendre transforming this Hamiltonian and reintroducing $v^0$, one 
obtains the Lagrangian given in Appendix \ref{su2comp}, which gives 
the SU(2) modified Gauss law
\be \label{gauss}
{\cal D}_i \Pi^{i g} = - \eps^{gac} {\cal I}^{ab}
(A^c {\cal D}^0 A^{\dag b} + A^{\dag c} {\cal D}^0 A^b 
+ i \psi^b \sigma^0 \bpsi^c + i \lambda^b \sigma^0 \blambda^c)
\ee

\noindent
We now come to one of the main purposes of our paper, which is to
compute the central charge for the SU(2) theory, 
$Z=\frac{i}{4}\epsilon^{\alpha \beta} \{ Q_{1 \alpha} , Q_{2 \beta} \}_{+}$.
After some fairly straightforward computations we obtain
\bea
Z &=& \int d^3 x {\Big (} \partial_i  
[i \sqrt2 (\Pi^{a i} A^{\dag a} + B^{a i} {\cal F}^{\dag a})
- {\cal F}^{\dag ab} \bar\psi^a \bar\sigma^{i 0} \bar\lambda^b] \nonumber \\
&+& i\sqrt2 [(D_i \Pi^{ai}) A^{\dagger a}  
+ i {\cal I}^{be} \epsilon^{bcd} A^{\dagger d} 
(\psi^e \sigma^0 \bar\psi^c + \lambda^e \sigma^0 \bar\lambda^c) 
-  \epsilon^{abc} A^b A^{\dagger c} \pi^a_A]  {\Big )}
\eea
where the properties of ${\cal F}^{a \cdots b}$ listed in the Appendix 
\ref{su2comp} were extensively used. 
We see from here that the terms which are not a total divergence, given 
in the second line above, simply cancel due to the Gauss law (\ref{gauss}).
Eventually we are left with the surface terms that vanish when the SU(2) 
gauge symmetry is not broken down to U(1). If we break the symmetry along a 
flat direction of the Higgs potential, say $a=3$, we recover the same result
we found in the U(1) sector. In other words we see that on the sphere at
infinity
\bea
Z &=& \int d^2 \vec{\Sigma} \cdot 
[i \sqrt2 (\vec{\Pi^a} A^{\dag a} + \frac{1}{4\pi} \vec{B}^a 
{\cal F}^{\dag a}) 
- \frac{1}{4\pi} {\cal F}^{\dag ab} \bpsi^a\vec{\bsigma}\blambda^b)] 
\nonumber \\
&\to& i \sqrt2 \int d^2 \vec{\Sigma} \cdot 
(\vec{\Pi^3} A^{\dag 3} + \frac{1}{4\pi} \vec{B}^3 A^{\dag 3}_D)
\label{zsu2eff}
\eea
where $\vec{\bsigma} \equiv (\bsigma^{01}, \bsigma^{02}, \bsigma^{03})$,
$\partial {\cal F}^{\dag} / \partial A^{\dag 3} = A^{\dag 3}_D$, and we 
reintroduced the factor $4\pi$. 
We also made the usual assumption that the bosonic massive fields in the 
SU(2)/U(1) sector ($a=1,2$) and all the fermionic fields fall off faster than 
$r^3$, whereas the scalar massless field ($a=3$) and its dual tend to their 
Higgs v.e.v.'s $a^*$ and $a^*_D$, respectively.

\noindent
Note that, in the broken phase, the formula (\ref{zsu2eff}) is the same as 
the U(1) formula (\ref{zeffu1}). Thus we conclude that the fields in the 
massive sector have no effect on the mass formula.

\noindent
{\bf Acknowledgements.}

\noindent
The authors would like to thank M.Magro, I.Sachs and F.Delduc for useful 
discussions.
A.I. was partially supported by the fellowship ``Theoretical Physics of the 
Fundamental Interactions, COFIN 1999'' - University of Rome Tor Vergata.
S.W. was supported by the Swiss National Science Foundation during his staying
at the Ecole Normale Sup\'erieure de Lyon. 

\newpage

\noindent
{\bf \huge Appendices}

\appendix

\section{Notation and Conventions}

\noindent
In this Appendix we explain the notation and conventions used for the spinors,
the Poisson brackets and the SU(2) gauge group.
 
\subsection{Spinor Conventions and Useful Algebra}\label{not}

\noindent
We follow Wess and Bagger \cite{wb} without changes. 

\noindent
The spinors are Weyl two components in Van der Waerden notation.
Spinors with un-dotted indices transform under the representation
$(\frac{1}{2},0)$ of $SL(2,\C)$ and spinors with dotted indices
transform under the conjugate representation $(0,\frac{1}{2})$.
The metric is $\eta_{\mu \nu} = {\rm diag}(-1,1,1,1)$. 
To rise and lower the spinor indices 
we use $\eps_{\alpha \beta}$ and $\eps^{\alpha \beta}$, where 
$\eps_{2 1} = \eps^{1 2} = - \eps_{1 2} = - \eps^{2 1} = 1$. Also
$\eps_{0 1 2 3} = -1$.
To raise and lower spinor indices use A(9) in \cite{wb} always matching
the indices from left to right: 
$\psi^\alpha = \eps^{\alpha \beta} \psi_\beta$ and 
$\psi_\alpha = \epsilon_{\alpha \beta} \psi^\beta$.
Note that momenta are on a different footing since they are derivatives of the 
Lagrangian w.r.t. a grassmanian field. Thus the indices have to be raised and 
lowered with the opposite convention 
$
\pi^\alpha = \epsilon^{\beta \alpha} \pi_\beta
$
and
$
\pi_\alpha = \epsilon_{\beta \alpha} \pi^\beta
$.
Note that $\psi \chi = \chi \psi$ ($\bar\psi \bar\chi = \bar\chi \bar\psi$)
but $\pi \chi = - \chi \pi$ ($\bar\pi \bar\chi = - \bar\chi \bar\pi$). 
Explicitly writing the indices that means: 
$\pi^\alpha \chi_\alpha = \pi_\alpha \chi^\alpha$
and $\bar\pi_{\dot \alpha} \bar\chi^{\dot \alpha} = 
\bar\pi^{\dot \alpha} \bar\chi_{\dot \alpha}$.

\noindent
Beside the identities given in \cite{wb} we also find 
\bea
\psi^\alpha \lambda^\beta - \psi^\beta \lambda^\alpha = 
- \eps^{\alpha \beta} \psi\lambda
&,&
\psi_\alpha \lambda_\beta - \psi_\beta \lambda_\alpha = 
\eps_{\alpha \beta} \psi\lambda \\
\bpsi^{\dalpha} \blambda^{\dbeta} - 
\bpsi^{\dbeta} \blambda^{\dalpha} =
\eps^{\dalpha \dbeta} \bpsi \blambda
&,&
\bpsi_{\dalpha} \blambda_{\dbeta} - 
\bpsi_{\dbeta} \blambda_{\dalpha} = 
- \eps_{\dalpha \dbeta} \bpsi \blambda 
\eea
and the following identities for the $\sigma$ matrices with free spinor indices
\bea
\sigma^{\mu \nu \: \beta}_{\: \alpha} \sigma_{\nu \: \gamma \dot\gamma} &=&
\frac{1}{2} (\sigma^\mu_{\: \delta \dot\gamma}
\epsilon_{\gamma \alpha} \epsilon^{\beta \delta}  -
\sigma^\mu_{\: \alpha \dot\gamma} \delta^\beta_\gamma) \label{sigsig1} \\
\sigma^{\mu \nu \: \beta}_{\: \alpha} \bar\sigma^{\dot\alpha \gamma}_{\nu} &=& 
\frac{1}{2} (\bar\sigma^{\mu \: \dot\alpha \delta} 
\epsilon_{\alpha \delta} \epsilon^{\beta \gamma} +
\bar\sigma^{\mu \: \dot\alpha \beta} \delta_\alpha^\gamma ) \label{sigsig2} \\
\bar\sigma^{\mu \nu \: \dot\alpha}_{\: \quad \dot\beta}
\bar\sigma^{\dot\gamma \gamma}_{\nu} &=& 
\frac{1}{2} (\bar\sigma^{\mu \: \dot\delta \gamma}
\epsilon^{\dot\alpha \dot\gamma}\epsilon_{\dot\delta \dot\beta} -
\bar\sigma^{\mu \: \dot\alpha \gamma} 
\delta^{\dot\gamma}_{\dot\beta}) \label{sigsig3} \\
\bar\sigma^{\mu \nu \: \dot\alpha}_{\: \quad \dot\beta}
\sigma_{\nu \: \alpha \dot\gamma} &=& \frac{1}{2}
(\sigma^\mu_{\: \alpha \dot\delta}
\epsilon^{\dot\alpha \dot\delta}\epsilon_{\dot\beta \dot\gamma} +
\sigma^\mu_{\: \alpha \dot\beta} \delta^{\dot\alpha}_{\dot\gamma}) 
\label{sigsig4}
\eea
that imply
\be
\sigma^{\mu \nu} \sigma_{\nu} = 
\sigma_\nu \bar\sigma^{\nu \mu} =
- \frac{3}{2} \sigma^\mu
\quad \quad
\bar\sigma^{\mu \nu} \bar\sigma_{\nu} = 
\bar\sigma_\nu \sigma^{\nu \mu} = 
- \frac{3}{2} \bar\sigma^\mu
\ee
We also have 
\bea
\sigma^{\mu \nu \beta}_\alpha \bsigma_{\mu \nu \dbeta}^{\dalpha}
&=& - \delta^{\dalpha}_{\dbeta} \delta_\alpha^\beta \\
\sigma^{0 \mu \beta}_\alpha \sigma_{\gamma 0 \mu}^\delta
&=& - \frac{1}{4} (\eps_{\alpha \gamma} \eps^{\beta \delta} 
+ \delta^\delta_\alpha \delta^\beta_\gamma) \\
\bsigma^{0 \mu \dalpha}_{\dot\beta} \bsigma^{\dot\gamma}_{0 \mu \dot\delta}
&=& - \frac{1}{4} (\eps^{\dalpha \dot\gamma} \eps_{\dot\beta \dot\delta} 
+ \delta_{\dot\delta}^{\dot\alpha} \delta_{\dbeta}^{\dot\gamma})
\eea
Also useful are the identities:
\be
(\sigma^{\rho \sigma} \sigma^{\mu \nu})^{\alpha}_{\beta}
v_{\rho \sigma}v_{\mu \nu} = 
- \frac{1}{2} \delta^{\alpha}_{\beta} v_{\mu \nu} {\hat v}^{\mu \nu}
\quad \quad 
(\bar\sigma^{\rho \sigma} \bar\sigma^{\mu \nu})^{\dot\alpha}_{\dot\beta}
v_{\rho \sigma}v_{\mu \nu} = 
- \frac{1}{2} \delta^{\dot\alpha}_{\dot\beta} v_{\mu \nu} 
{\hat v}^{\dagger \mu \nu}
\ee
\bea
\sigma^{\rho \sigma}\sigma^{\mu}v_{\rho \sigma} &=& 
{\hat v}^{\mu \nu} \sigma_\nu  \nonumber \\
\bar\sigma^{\mu} \sigma^{\rho \sigma}v_{\rho \sigma} &=& 
- {\hat v}^{\mu \nu} \bar\sigma_\nu  \\
\sigma^{\mu} \bar\sigma^{\rho \sigma} v_{\rho \sigma} &=& 
- {\hat v}^{\dagger \mu \nu} \sigma_\nu \nonumber \\
\bar\sigma^{\rho \sigma} \bar\sigma^{\mu} v_{\rho \sigma} &=& 
{\hat v}^{\dagger \mu \nu} \bar\sigma_\nu \nonumber 
\eea
where 
\be
{\hat v}^{\mu \nu} = v^{\mu \nu} + \frac{i}{2} v^{* \mu \nu}
\quad \quad
{\hat v}^{\dagger \mu \nu} = v^{\mu \nu} - \frac{i}{2} v^{* \mu \nu}
\ee
and $v^{\mu \nu} = - v^{\nu \mu}$.

\subsection{Graded Poisson Brackets}\label{poisson}

\noindent
We deal with c-number valued fields, i.e. non operator, in the classical 
as well as effective case. Therefore the Susy algebra has to be implemented 
via graded Poisson brackets, namely with Poisson brackets $\{ , \}_{-}$ and 
anti-brackets $\{ , \}_{+}$. We define the following equal time Poisson (anti) 
brackets\footnote{Following a nice argument given by Dirac \cite{dir}, this 
definition leads to the quantum {\it anti}-commutator for two fermions. The 
original argument relates classical Poisson brackets to the commutator 
$[B_1(x) , B_2(y)]_{-} \to i \hbar \{ B_1(x) , B_2(y)\}_{-}$, where the $B$'s
stand for bosonic variables. The generalization to fermions is naturally 
given by $ [F_1(x) , F_2(y)]_{+} \to i \hbar \{ F_1(x) , F_2(y)\}_{+}$
(we shall use the natural units $\hbar = c = 1$), where the $F$'s are 
fermionic variables. See also \cite{gara}.}
\bea
\{ B_1(x) , B_2(y)\}_{-}&\equiv& \int d^3 z 
 {\bigg (} \frac{\delta B_1(x)}{\delta \Phi(z)} 
          \frac{\delta B_2(y)}{\delta \Pi(z)} - 
	  \frac{\delta B_2(y)}{\delta \Phi(z)}
	  \frac{\delta B_1(x)}{\delta \Pi(z)} {\bigg )} \\
\{ B(x) , F(y)\}_{-} &\equiv& \int d^3 z 
 {\bigg (} \frac{\delta B(x)}{\delta \Phi(z)} 
          \frac{\delta F(y)}{\delta \Pi(z)} - 
	  \frac{\delta F(y)}{\delta \Phi(z)}
	  \frac{\delta B(x)}{\delta \Pi(z)} {\bigg )} \\
\{ F_1(x) , F_2(y)\}_{+} &\equiv& \int d^3 z 
 {\bigg (} \frac{\delta F_1(x)}{\delta \Phi(z)} 
          \frac{\delta F_2(y)}{\delta \Pi(z)} + 
	  \frac{\delta F_2(y)}{\delta \Phi(z)}
	  \frac{\delta F_1(x)}{\delta \Pi(z)} {\bigg )} 
\eea
where the $B$'s are bosonic and the $F$'s fermionic variables and $\Phi$
and $\Pi$ span the whole phase space. 

\noindent
Form this definition it follows that the properties of the graded Poisson
brackets are the same as for standard commutators and anti-commutators
\be
\{ B_1 , B_2 \}_{-} = - \{ B_2 , B_1 \}_{-} \quad
\{ B , F \}_{-} = - \{ F , B \}_{-} \quad
\{ F_1 , F_2 \}_{+} = + \{ F_2 , F_1 \}_{+}
\ee
thus
\bea
\{ B_1 \; , \; B_2 B_3 \}_{-} &=& \{ B_1 , B_2 \}_{-} B_3 
+ B_2 \{ B_1 , B_3 \}_{-} \\
\{ B_1  B_2 \; , \; B_3 \}_{-} &=& B_1 \{ B_2 , B_3 \}_{-}
+ \{ B_1 , B_3 \}_{-} B_2 
\eea
\bea
\{ F_1 \; , \; F_2 F_3 \}_{-} &=& \{ F_1 , F_2 \}_{+} F_3 
- F_2 \{ F_1 , F_3 \}_{+} \\
\{ F_1  F_2 \; , \; F_3 \}_{-} &=& F_1 \{ F_2 , F_3 \}_{+}
- \{ F_1 , F_3 \}_{+} F_2 
\eea
\bea
\{ F_1 F_2 \; , \; B \}_{-} &=& F_1 \{ F_2 , B \}_{-} 
+  \{ F_1 , B \}_{-} F_2 \\
\{ B \; , \; F_3 F_4 \}_{-} &=&  \{ B , F_3 \}_{-} F_4
+ F_3 \{ B , F_4 \}_{-} 
\eea
\bea
\{ F_1 F_2 \; , \; F_3 F_4 \}_{-} &=& F_1 \{ F_2 , F_3 \}_{+} F_4 
- F_1 \{ F_2 , F_4 \}_{+} F_3 \\
&& - F_2 \{ F_1 , F_3 \}_{+} F_4  
+ F_2 \{ F_1 , F_4 \}_{+} F_3 
\eea
\noindent
Let us notice that only a formal algebraic meaning can be associated to the 
Poisson anti-bracket of two fermions, since there is no physical meaning for 
a {\it classical} fermion.

\noindent
The canonical equal-time Poisson brackets for a Lagrangian with bosonic and 
fermionic fields are given by the usual expressions. The same structure 
survives at the effective level even if a great deal of care is required. For 
instance the non-zero equal-time Poisson brackets for the U(1) sector of the 
effective SW theory are
\be
\{A(x), \pi_A (y) \}_{-} = \{A^\dag (x), \pi_{A^\dag} (y) \}_{-} =
\delta^{(3)} (\vec{x} - \vec{y}) \quad , \quad
\{v_i (x), \Pi^j (y) \}_{-} = \delta_i^j \delta^{(3)} (\vec{x} - \vec{y})
\ee
and
\be 
\{ \bar\psi_{\dot\alpha} (x) \;,\; (\pi^{I}_{\bar \psi})^{\dot\beta} (y) \}_{+}
= \{ \bar\lambda_{\dot\alpha} (x) \;,\; 
(\pi^{I}_{\bar \lambda})^{\dot\beta} (y) \}_{+} =
\delta_{\dot\alpha}^{\dot\beta} \delta^{(3)} (\vec{x} - \vec{y}) 
\label{canfermiI}
\ee
or
\be
\{ \psi_{\alpha} (x) \;,\; (\pi^{II}_{\psi})^{\beta} (y) \}_{+}
= \{ \lambda_{\alpha} (x) \;,\; (\pi^{II}_{\lambda})^{\beta} (y) \}_{+} =
\delta_{\alpha}^{\beta} \delta^{(3)} (\vec{x} - \vec{y}) 
\label{canfermiII} 
\ee
depending on the phase space one chooses. Otherwise, in both cases, one can 
write
\be
\{\psi_{\alpha}(x) \;,\; \bar\psi_{\dot\alpha} (y)\}_{+} 
= - \frac{i}{\cal I} \sigma^0_{\alpha \dot\alpha} 
\delta^{(3)} (\vec{x} - \vec{y})  \label{fermi} 
\ee
(same for $\lambda$).
\noindent
Note that $\{ \Pi^i \;,\; \chi \}_{-} = 0$, where 
$\chi \equiv \psi, \lambda, \bpsi, \blambda$, even if the effective $\Pi^i$
contains all the fermions of the theory. Finally, we have that 
$
\{ \pi_A \;,\; {\cal I} \}_{-} = - \frac{1}{2i} {\cal F}^{(3)}
$
and 
$
\{ \pi_{A^\dag} \;,\; {\cal I} \}_{-} = \frac{1}{2i} {{\cal F}^{(3)}}^\dag
$
thus the usual assumption that fermions and bosons should commute no longer 
holds
\bea
\{ \pi_A \;,\; \pi_{\bpsi} \}_{-} = 0 &\Rightarrow&
\{ \pi_A \;,\; \psi \}_{-} = -\frac{i}{2} f \psi \\
\{ \pi_{A^\dag} \;,\; \pi_{\bpsi} \}_{-} = 0 &\Rightarrow&
\{ \pi_{A^\dag} \;,\; \psi \}_{-} = + \frac{i}{2} f^\dag \psi
\eea
with $f = {\cal F}^{(3)} / {\cal I}$ (same for $\lambda$). 

\noindent 
Note that  
$
\{ \psi_\alpha \; , \; \pi_\psi^\beta \}_{+} = \delta_\alpha^\beta
$
and
$
\{ \psi^\alpha \; , \; \pi_{\psi \beta} \}_{+} = \delta^\alpha_\beta
$
are compatible iff $\pi_\alpha = - \eps_{\alpha \beta} \pi^\beta$.
Note also that we impose 
$
\{ \psi_\alpha \; , \; \pi_{\psi \beta} \}_{+} = \eps_{\alpha \beta}
=  \{ \pi_{\psi \alpha} \; , \;  \psi_\beta \}_{+} 
$.

\subsection{SU(2) Conventions}\label{su2conv}

\noindent
A generic SU(2) vector is defined as $\vec{X} = \frac{1}{2} \sigma^a X^a$ with
$a=1,2,3$ and we follow the summation convention. The $\sigma^a$'s are
the standard Pauli matrices satisfying 
$[\sigma^a, \sigma^b] = 2 i \eps^{a b c} \sigma^c$, where $\eps^{a b c}$ are 
the structure constants of SU(2), and 
${\rm Tr} \sigma^a \sigma^b = 2 \delta^{a b}$. The covariant derivative 
and the vector field strength are given by
${\cal D}_\mu \vec{X} = \partial_\mu \vec{X} - i [\vec{v}_\mu , \vec{X}]$ 
and $\vec{v}_{\mu \nu}=\partial_\mu \vec{v}_\nu - \partial_\nu \vec{v}_\mu 
- i [\vec{v}_\mu , \vec{v}_\nu]$, respectively.

\noindent
Some authors keep the renormalizable 
SU(2) gauge coupling $g$ even in the effective theory (for instance their 
covariant derivatives are defined as 
${\cal D}_\mu X^a = \partial_\mu X^a + g \eps^{a b c} v^b_\mu X^c$). 
This is somehow misleading since in SW theory the 
effective coupling is once and for all given by $\tau(a) = {\cal F}^{(2)}(a)$.
Of course the microscopic theory is scale invariant before SSB\footnote{As a 
matter of fact, it is invariant under the full superconformal group.}, and a 
redefinition of the fields $g X \to X$ does no harm. The matter is less clear 
in the effective theory, where even the definition of what is a field poses 
some problems and scale invariance is lost after SSB. Therefore we prefer to 
follow the conventions of \cite{sw}, where already at 
microscopic level the $g$ is absorbed in the definition of the fields and 
only appears in the overall factor $1 / g^2$ (see also our expression for 
the U(1) classical Lagrangian in (\ref{lu1class})).

\noindent
Nevertheless we can keep track of $g$ since by charge conjugation
$g \to -g$ (see for instance \cite{iz}), which in our notation becomes 
$\eps^{abc} \to -\eps^{abc}$.

\noindent

\noindent
The SU(2) transformations for the first Susy are given by 
\bea
\delta_1 \vec{A} &=& \sqrt2 \eps_1 \vec{\psi} \nonumber \\
\delta_1 \vec{\psi}^\alpha &=& \sqrt2 \eps_1^\alpha \vec{E}\\
\delta_1 \vec{E}  &=& 0 \nonumber 
\eea
\bea
\delta_1 \vec{E}^\dag &=& i \sqrt2 \eps_1 \not\!{\cal D} \vec{\bpsi}
+ 2 i [\vec{A}^\dag , \eps_1 \vec{\lambda}] 
\nonumber \\
\delta_1 \vec{\bpsi}_{\dalpha} &=& - i \sqrt2 \eps_1^\alpha 
\not\!{\cal D}_{\alpha \dalpha} \vec{A}^\dag \\
\delta_1 \vec{A}^\dag &=& 0 \nonumber 
\eea
\bea
\delta_1 \vec{\lambda}^\alpha &=& 
- \eps^\beta_1 ( \sigma^{\mu \nu \: \alpha}_{\: \beta} \vec{v}_{\mu \nu}
    - i \delta^\alpha_\beta \vec{D} ) \nonumber \\
\delta_1 \vec{v}^\mu &=& i \eps_1 \sigma^\mu \vec{\blambda} \quad , \quad
\delta_1 \vec{D} = - \eps_1 \not\!{\cal D} \vec{\blambda} \\
\delta_1 \vec{\blambda}_{\dalpha} &=& 0 \nonumber 
\eea

\noindent
The SU(2) effective canonical momenta are given by
\be
\Pi^{a i} = -\frac{1}{2i} 
({\cal F}^{a b} {\hat v}^{0 i b} - 
{\cal F}^{a b \dagger} {\hat v}^{\dagger 0 i b})
+ \frac{1}{i \sqrt2} {\cal F}^{a b c} \lambda^b \sigma^{0 i} \psi^c 
- \frac{1}{i \sqrt2} {\cal F}^{a b c \dagger} 
     \bar\lambda^b \bar\sigma^{0 i} \bar\psi^c \label{Pisu2}
\ee
\be
\pi^a_A = - {\cal I}^{a b} \partial^0 A^{\dagger b} \quad , \quad
\pi^a_{A^\dagger} = - {\cal I}^{a b} \partial^0 A^b 
- \frac{1}{2} {{\cal F}^\dagger}^{a b c} (\bar\psi^b \bar\sigma^0 \psi^c
+ \bar\lambda^b \bar\sigma^0 \lambda^c) \label{pbsu2}
\ee
\be
\pi^a_{\bar \psi} = i {\cal I}^{a b} \psi^b \sigma^0 
\quad , \quad
\pi^a_{\bar \lambda} = i {\cal I}^{a b} \lambda^b  \sigma^0 \label{pfsu2}
\ee
where we choose the {\it setting $I$} (see the correspondent U(1) expressions) 
and the temporal gauge for the vector field (thus ${\cal D}^0 = \partial^0$).
Their classical limit is obtained by simply using 
${\cal F}^{a b} \to \tau \delta^{a b}$ and ${\cal F}^{a b \cdots c} \to 0$.

\noindent
Finally, the SU(2) prepotential $\cal F$ is a holomorphic function of the 
SU(2) gauge Casimir $A^a A^a$, $a = 1,2,3$. Our $\cal F$ corresponds to the 
function $\cal H$ in Seiberg and Witten conventions \cite{sw}.
Some care is necessary in handling its derivatives, the first 
four being given by
\bea
{\cal F}^a &=& 2 A^a {\cal F}^{(1)} \label{Fder1}\\
{\cal F}^{ab} &=& 2 \delta^{ab}{\cal F}^{(1)}+ 4 A^a A^b {\cal F}^{(2)} \label{Fder2}\\
{\cal F}^{abc} &=& 4 (\delta^{ab} A^c + \delta^{ac} A^b +\delta^{bc} A^a)
{\cal F}^{(2)}+ 8 A^a A^b A^c {\cal F}^{(3)} \label{Fder3} \\
{\cal F}^{abcd} &=& 4 {\cal F}^{(2)}
(\delta^{ab} \delta^{cd} + \delta^{ac} \delta^{bd} +\delta^{bc} \delta^{ad})
+ 8 {\cal F}^{(3)}(A^a A^b \delta^{cd} + A^a A^c \delta^{bd} 
+ A^a A^d \delta^{bc} \nonumber \\
&& + A^b A^c \delta^{ad} + A^b A^d \delta^{ac} 
+ A^c A^d \delta^{ab}) + 16 {\cal F}^{(4)} A^a A^b  A^c A^d \label{Fder4}
\eea
similarly for ${\cal F}^\dag$.

\noindent
Form the expressions (\ref{Fder1})-(\ref{Fder4}) it is easy to obtain the 
following very useful identities: 
\bea
\epsilon^{abc} {\cal F}^{bd} A^c &=& \epsilon^{adc} {\cal F}^c \label{F1} \\ 
\epsilon^{bcd} {\cal F}^{be} A^d &=& - \epsilon^{bed} {\cal F}^{bc} 
A^d \label{F2} \\
{\cal F}^{abc} \epsilon^{cde} A^e &=& 
{\cal F}^{be} \epsilon^{ade} + {\cal F}^{ae} \epsilon^{bde} \label{F3}
\eea
similarly for ${\cal F}^\dag$.

\noindent
Properties (\ref{F1})-(\ref{F3}) are extensively used throughout the SU(2)
computations. 

\section{Computation of $V^\mu$ for the U(1) Theory}\label{compV}

\noindent
We first notice that, by varying off-shell the Lagrangian (\ref{lu1eff}) 
under the Susy transformations given in (\ref{trns1})-(\ref{trns2}), there is 
no mixing of the $\cal F$ terms with the ${\cal F}^\dag$ terms. The structure 
of this Lagrangian is 
\bea
2 i {\cal L} &=&  - {\cal F}^{(2)} [2B+2F] + {\cal F}^{(3)} [1B 2F] 
+ {\cal F}^{(4)} [4F] \\
&+& {{\cal F}^{(2)}}^\dag [2B+2F]^\dag - {{\cal F}^{(3)}}^\dag [1B 2F]^\dag
- {{\cal F}^{(4)}}^\dag [4F]^\dag 
\eea
where $B$ and $F$ stand for bosonic and fermionic variables, respectively.

\noindent
For instance, if we vary the $\cal F$ terms under $\delta_1$ we have  
$(\delta_1 {\cal F}^{(4)}) [3] = 0$, whereas the other terms combine as follows
\bea
{\cal F}^{(4)} \delta_1 [4F] \sim {\cal F}^{(4)} (1B 3F) 
&{\rm with}&
(\delta_1 {\cal F}^{(3)}) [1B 2F] \sim {\cal F}^{(4)} (1B 3F) \nonumber \\
{\cal F}^{(3)} \delta_1 [1B 2F] \sim {\cal F}^{(3)} (2B 1F + 3F) 
&{\rm with}&
(\delta_1 {\cal F}^{(2)}) [2B+2F] \sim {\cal F}^{(3)} (2B 1F + 3F) \nonumber
\eea
Finally there are the terms ${\cal F}^{(2)} \delta_1 [2B + 2F]$, which are the
naive generalization of the classical $V_1^\mu$. The aim is to write these 
quantities as one single total divergence and express it in terms of momenta 
and variations of the fields.

\noindent
This computation is by no means easy. It is matter of
\begin{itemize}
\item{identifying similar terms and compare them}
\item{use partial integration cleverly: never throw away surface terms!}
\item{use extensively Fierz identities and spinor algebra}
\end{itemize}
What we shall show explicitly in this Appendix, is only the simplest part 
of the computation of $V^\mu_1$, namely the contribution coming from the 
$\cal F$-terms. 

\noindent
Let us apply the scheme discussed above. First we consider the 
${\cal F}^{(4)}$ type of terms. If we find contributions from these terms we 
know that they cannot be cancelled by terms coming from the rigid current 
$N_\mu$ and there is no hope to rearrange them in the form of on-shell 
dummy fields (they only contain ${\cal F}^{(3)}$ type of terms). This would 
then be a signal that by commuting the charges we could have contributions that
would spoil the SW mass formula. What we find is that the terms
\be
{\cal F}^{(4)} \delta_1 [4F] =
{\cal F}^{(4)} \frac{1}{2}[(\delta_1\psi)\psi \lambda^2  
+ \psi^2 (\delta_1\lambda)\lambda] 
={\cal F}^{(4)} \frac{1}{2}[\sqrt2 \eps_1\psi \lambda^2 E  
-\eps_1 \sigma^{\mu \nu}\lambda v_{\mu \nu} \psi^2
+ i \eps_1\lambda \psi^2 D]
\ee
added to the terms
\be
(\delta_1 {\cal F}^{(3)}) [1B 2F] = 
{\cal F}^{(4)} [\eps_1\psi \lambda\sigma^{\mu \nu}\psi v_{\mu \nu}
-\frac{1}{\sqrt2} E \eps_1\psi \lambda^2 
+i D \eps_1\psi \psi\lambda]
\ee
fortunately give zero.

\noindent
Let us then move to the next level, the ${\cal F}^{(3)}$ terms. In principle 
these terms can be present, since they appear in the expression of the 
on-shell dummy fields. We find that the terms
\bea
{\cal F}^{(3)}  [(\delta_1 1B) 2F] &=& {\cal F}^{(3)}
[\frac{1}{\sqrt 2} \lambda \sigma^{\mu \nu} \psi \delta_1 v_{\mu \nu}
-\frac{1}{2} (\delta_1 E^\dag) \psi\psi 
+\frac{i}{\sqrt 2} (\delta_1 D) \psi\lambda] \nonumber \\
&=& {\cal F}^{(3)}
[\frac{1}{\sqrt 2} \lambda \sigma^{\mu\nu} \psi 
2 i \eps_1 \sigma_\nu \partial_\mu \blambda
-\frac{i}{\sqrt2} \eps_1\not\!\partial\bpsi \psi\psi 
-\frac{i}{\sqrt 2} \eps_1\not\!\partial\blambda  \psi\lambda] \nonumber \\
&=& \frac{i}{\sqrt2} {\cal F}^{(3)} 
(2 \lambda\not\!\partial\blambda  \eps_1\psi
- \eps_1\not\!\partial\bpsi \psi\psi)
\eea
summed to the terms
\be
-(\delta_1 {\cal F}^{(2)}) [2F] = -i \sqrt2 {\cal F}^{(3)} 
\lambda\not\!\partial\blambda  \eps_1\psi
-i \sqrt2 {\cal F}^{(3)} \psi\not\!\partial\bpsi \eps_1\psi)
\ee
again give zero. 

\noindent
What is left are the other ${\cal F}^{(3)}$ terms and the ${\cal F}^{(2)}$ 
terms. There we find
\bea
-{\cal F}^{(2)} \delta_1 [2B+2F] &=& 
-{\cal F}^{(2)} [ (\partial_\mu A^\dagger) \partial^\mu (\delta_1 A)
+ \frac{1}{2} (\delta_1 v_{\mu \nu}) v^{\mu \nu}
+ \frac{i}{4} (\delta_1 v_{\mu \nu}) v^{* \mu \nu} \nonumber \\
&& + i (\delta_1 \psi)  \not\!\partial {\bar \psi} 
+ i \psi \not\!\partial (\delta_1 {\bar \psi}) 
+ i (\delta_1 \lambda)  \not\!\partial {\bar \lambda}
- E \delta_1 E^\dag - D \delta_1 D] \nonumber \\
&=& -{\cal F}^{(2)} [ (\partial_\mu A^\dagger) \partial^\mu (\sqrt2 \eps_1\psi)
+ (i \eps_1 \sigma_\nu (\partial_\mu \blambda)) v^{\mu \nu}
\nonumber \\
&& - \frac{1}{2} (i \eps_1 \sigma_\nu (\partial_\mu \blambda)) v^{* \mu \nu} 
+ i \sqrt2 \eps_1 \not\!\partial {\bar \psi} E 
- \sqrt2 \psi^\alpha (\not\!\partial_{\alpha \dalpha}
\not\!\partial^{\dalpha \beta} A^\dag) \eps_{1 \beta} 
\nonumber \\
&& - i \eps_1^\beta (\sigma^{\mu \nu \: \alpha}_{\: \beta} v_{\mu \nu}
- i\delta^\alpha_\beta D )
\not\!\partial_{\alpha \dalpha}{\blambda}^{\dalpha} \nonumber \\
&& - i \sqrt2 \eps_1 \not\!\partial {\bpsi}E  
- D \eps_1\not\!\partial {\blambda}] \nonumber \\
&=& -{\cal F}^{(2)} [ \sqrt2 \eps_1(\partial^\mu \psi) \partial_\mu A^\dag
+ \sqrt2 \eps_1\psi \Box A^\dag] \nonumber \\
&=& -{\cal F}^{(2)} \partial^\mu [\sqrt2 \eps_1\psi \partial_\mu A^\dag]
\label{1}
\eea
Thus we find the first non zero contribution. Let us note that this term 
would already be a total divergence if we impose the classical limit 
${\cal F}^{(2)} \to \tau$. Thus we can guess that the ${\cal F}^{(3)}$ terms 
have to combine to give the quantum piece missing in order to built up 
a total divergence when summed to the terms (\ref{1}). We find that
\bea
-\delta_1 {\cal F}^{(2)} [2B] &=& {\cal F}^{(3)} \sqrt2 \eps_1 \psi
[- \partial_\mu A^\dagger \partial^\mu A
- \frac{1}{4} v_{\mu \nu}{\hat v}^{\mu \nu} + E E^\dag + \frac{1}{2} D^2]
\nonumber
\eea
summed to
\bea
{\cal F}^{(3)}  [1B (\delta_1 2F)] &=& {\cal F}^{(3)}
[\frac{1}{\sqrt 2} (\delta_1\lambda) \sigma^{\mu \nu} \psi v_{\mu \nu}
+ \frac{1}{\sqrt 2} \lambda \sigma^{\mu \nu} (\delta_1 \psi) v_{\mu \nu} 
\nonumber \\
&& -\frac{1}{2} (E^\dag \psi\delta_1\psi + E \lambda\delta_1\lambda)
+\frac{i}{\sqrt 2} D (\delta_1 \psi) \lambda 
+\frac{i}{\sqrt 2} D \psi \delta_1 \lambda] \nonumber \\
&=& {\cal F}^{(3)}
[- \frac{1}{\sqrt 2} \eps_1 \sigma^{\mu \nu} \sigma^{\rho \sigma}\psi 
v_{\mu \nu} v_{\rho \sigma} 
+ \frac{i}{\sqrt 2} \eps_1 \sigma^{\mu \nu}\psi v_{\mu \nu} D \nonumber \\
&& + \lambda \sigma^{\mu \nu}\eps_1  v_{\mu \nu} E 
- E^\dag E \sqrt2 \eps_1\psi 
+ E \eps_1 \sigma^{\mu \nu}\lambda v_{\mu \nu})
\nonumber \\ 
&& -i \eps_1\lambda DE + i D E \eps_1\lambda 
- \frac{i}{\sqrt 2} D \eps_1\sigma^{\mu \nu} v_{\mu \nu}\psi  
- \frac{1}{\sqrt 2} D^2 \eps_1\psi ] \nonumber \\
&=& - {\cal F}^{(3)} 
[\frac{1}{\sqrt 2} \eps_1 \sigma^{\mu \nu} \sigma^{\rho \sigma} \psi
v_{\mu \nu} v_{\rho \sigma}
+ E^\dag E \sqrt2 \eps_1\psi + \frac{1}{\sqrt 2} D^2 \eps_1\psi] \nonumber 
\eea
give
\be
( \partial^\mu {\cal F}^{(2)}) [-\sqrt2 \eps_1\psi \partial_\mu A^\dag]
\label{2}
\ee
Collecting the two contributions (\ref{1}) and (\ref{2}) we end up with 
the wanted total divergence
\be
{\cal F}^{(2)} \partial^\mu [-\sqrt2 \eps_1\psi \partial_\mu A^\dag] +
(\partial^\mu {\cal F}^{(2)}) [-\sqrt2 \eps_1\psi \partial_\mu A^\dag] 
= \partial^\mu (-{\cal F}^{(2)} \sqrt2 \eps_1\psi \partial_\mu A^\dag) 
= \partial^\mu (\frac{{\cal F}^{(2)}}{\cal I} \delta_1A \pi^\mu_A)
\label{3}
\ee
where the definitions of momenta and the Susy transformations were used.

\noindent
More labour is needed for the ${\cal F}^\dag$ terms. We only give the result 
of that computation here. We have 
\bea
\partial_\mu [{{\cal F}^{(2)}}^\dag \sqrt2 \eps_1\psi \partial^\mu A^\dag
+ i {{\cal F}^{(2)}}^\dag \eps_1 \sigma_\nu \blambda \hat{v}^{\mu \nu \dag}
+ {{\cal F}^{(2)}}^\dag \eps_1 \sigma^\mu \blambda D \nonumber \\
+ {{\cal F}^{(2)}}^\dag \sqrt2 \eps_1\sigma^\nu\sigma^\mu\psi \partial_\nu A^\dag
+ i \sqrt2 {{\cal F}^{(2)}}^\dag \bpsi \bsigma^\mu \eps_1 E
+ \frac{i}{\sqrt2}{{\cal F}^{(3)}}^\dag \eps_1 \sigma^\mu\bpsi \lambda^2]
\nonumber
\eea
Using the definitions of the non canonical momenta given in 
(\ref{ncm1})-(\ref{ncm2}) and the Susy transformations of the fields given in 
(\ref{trns1})-(\ref{trns2}), these terms can be recast into the following form
\be
\partial_\mu [-\frac{{{\cal F}^{(2)}}^\dag}{\cal I} \delta_1 A \pi^\mu_A
+ 2i\frac{{{\cal F}^{(2)}}^\dag}{{\cal F}^{(2)}} \delta_1 \bpsi \pi^\mu_{\bpsi}
+ 2i\delta_1\lambda \pi_\lambda^\mu 
+ {{\cal F}^{(2)}}^\dag \eps_1\sigma_\nu\blambda v^{* \mu \nu} 
+ 2 i \delta_1\psi \pi_\psi^\mu
+ \frac{i}{\sqrt2}{{{\cal F}^{(3)}}^\dag} \eps_1\sigma^\mu\bpsi \lambda^2]
\label{4}
\ee
Summing up the terms (\ref{3}) and (\ref{4}) and dividing by $2i$ we obtain
the final expression
\be
V_1^\mu = \delta_1 A \pi^\mu_A 
+  \frac{{{\cal F}^{(2) \dag}}}{{\cal F}^{(2)}} \delta_1 \bpsi \pi^\mu_{\bpsi} 
+ \delta_1 \psi \pi^\mu_\psi 
+ \delta_1 \lambda \pi^\mu_{\lambda} 
+ \frac{1}{2i} {{\cal F}^{(2) \dag}} \eps_1 \sigma_\nu \blambda v^{* \mu \nu}
+ \frac{1}{2\sqrt2} {{\cal F}^{(3) \dag}} \eps_1 \sigma^\mu \bpsi \blambda^2 
\ee
As explained earlier this form is not canonical and has to be modified 
according to the rules given there.

\section{U(1) ``Effective'' Susy Transformations}\label{vertrns}

\noindent
In this Section we want to explicitly prove that the U(1) effective charge 
given in (\ref{QI}) indeed generates the Susy transformations 
(\ref{trns1})-(\ref{trns2}). The component fields in the N=2 SYM multiplet
are $(A (A^\dag)$, $v^\mu, \psi (\bpsi)$, $\lambda (\blambda))$. 
We have: 
\bea
\Delta_1 A (x) \equiv \{ A (x) \; , \; \eps_1 Q_1^I \}_{-} &=& \int d^3 y 
\{ A (x) \; , \;  \delta_1 \bar{\psi} (y) 
\pi^{I}_{\bar\psi} (y) \}_{-} \nonumber \\
&=& \int d^3 y 
\{ A (x) \; , \; \sqrt2 \eps_1 \sigma^\nu \bar\sigma^0 \psi (y)
{\cal I} (y) \partial_\nu A^\dagger (y) \}_{-} \nonumber \\
&=& \int d^3 y
\{ A (x) \; , \; \sqrt2 \eps_1 \psi (y) \pi_A^I (y) + {\rm irr.} \}_{-}
\nonumber \\
&=& \sqrt2 \eps_1 \psi (x) = \delta_1 A (x) \label{d1A}
\eea
where ``irr.'' stands for terms irrelevant for the Poisson bracket.
\be
\Delta_1 A^\dagger (x) \equiv
\{ A^\dagger (x) \; , \; \eps_1 Q_1^I \}_{-} = 0 = \delta_1 A^\dagger (x)
\ee
\be
\Delta_1 v_i (x) \equiv \{ v_i (x) \; , \; \eps_1 Q_1^I \}_{-} = \int d^3 y 
\{ v_i (x) \; , \;  \Pi^j (y) \}_{-} \delta_1 v_j (y) 
= \delta_1 v_i (x)
\ee
\be
\Delta_1 \bpsi_{\dalpha} (x) \equiv 
\{\eps_1 Q_1^I \; , \;  \bar\psi_{\dot\alpha} (x) \}_{-} =
\int d^3 y  \delta_1 {\bar\psi}_{\dot\beta} (y) 
\{ \pi^{I \dot\beta}_{\bar\psi} (y) \;,\; \bar\psi_{\dot\alpha} (x) \}_{+}
=\delta_1 \bar{\psi}_{\dot\alpha} (x)
\ee
\be
\Delta_1 \bar\lambda_{\dot\alpha} (x) \equiv
\{\eps_1 Q_1^I  \; , \;  {\bar\lambda}_{\dot\alpha} (x) \}_{-} = 0 =
\delta_1 {\bar\lambda}_{\dot\alpha} (x)
\ee
For $\Delta_1 \pi^I_{\bar\psi \dot\alpha}$
some attention is due to the fact that $\pi^I_{\bar\psi \dot\alpha}$
is a product of a bosonic function $\cal I$ and of a fermion $\psi$. On the 
one hand
\be
\Delta_1 \pi^I_{\bar\psi \dot\alpha} (x) \equiv 
\{\eps_1 Q_1^I  \; , \;  \pi^I_{\bar\psi \dot\alpha} (x) \}_{-} 
= \int d^3 y ( - \frac{1}{2\sqrt2} {\cal F}^{(3) \dagger} 
\{ \epsilon_1 \sigma^0 \bar{\psi} \;,\; \pi^I_{\bar\psi \dot\alpha} (x) \}_{-}
\bar{\lambda}^2 ) 
= - \frac{1}{2\sqrt2} {\cal F}^{(3) \dagger} 
\eps_1^\alpha \sigma^0_{\alpha \dot\alpha} \bar{\lambda}^2
\ee
on the other hand,  writing explicitly $\pi^I_{\bar\psi \dot\alpha}$ we 
have 
\be
\Delta_1 \pi^I_{\bar\psi \dot\alpha} (x) =
\frac{1}{\sqrt2} {\cal F}^{(3)} \eps_1 \psi \psi^\alpha 
\sigma^0_{\alpha \dot\alpha} 
+ i {\cal I} \sigma^0_{\alpha \dot\alpha} \Delta_1 \psi^\alpha
\ee
where we have used 
$
\Delta_1 {\cal I} = \frac{1}{2i} 
({\cal F}^{(3)} \Delta_1 A - {\cal F}^{(3) \dag} \Delta_1 A^\dagger )
$.
Thus, by comparing the two expressions for 
$\Delta_1 \pi^I_{\bar\psi \dot\alpha}$ we obtain
\be
\Delta_1 \psi^\alpha = \sqrt2 \eps_1^\alpha 
(-\frac{i}{4} f \psi^2 + \frac{i}{4} f^\dagger {\bar\lambda}^2) 
= \sqrt2 \eps_1^\alpha E_{\rm on} = \delta_1^{\rm on} \psi^\alpha
\ee
where we have used the expression (\ref{dummy}) for $E$ on-shell and 
the Fierz identity 
$\psi_\alpha \psi^\beta = - \frac{1}{2} \delta_\alpha^\beta \psi^2$.

\noindent
More labour is needed to compute  $\Delta_1 \lambda$ from 
$\Delta_1 \pi^I_{\bar\lambda \dot\alpha}$. On the one hand
\bea
\Delta_1 \pi^I_{\bar\lambda \dot\alpha} (x) &\equiv&
\{\eps_1 Q_1  \; , \;  \pi^I_{\bar\lambda \dot\alpha} (x) \}_{-} \nonumber \\
&=& \int d^3 y (\Pi^i(y) \{ \delta_1 v_i(y) \;,\; 
\pi^I_{\bar\lambda \dot\alpha} (x) \}_{-} \nonumber \\
&& - \frac{1}{2i} {\cal F}^{(2) \dagger}(y) v^{* 0 i}(y) 
\{ \eps_1 \sigma_i \bar\lambda (y)
\;,\; \pi^I_{\bar\lambda \dot\alpha} (x) \}_{-} \nonumber \\
&& - \frac{1}{2\sqrt2} {\cal F}^{(3) \dagger}(y) \eps_1 \sigma^0 \bar\psi (y) 
\{ {\bar\lambda}^2 (y) \;,\; \pi^I_{\bar\lambda \dot\alpha} (x) \}_{-} )
\eea
On the other hand
\be
\Delta_1 \pi^I_{\bar\lambda \dot\alpha} =
\frac{1}{\sqrt2} {\cal F}^{(3)} \eps_1 \psi \lambda^\alpha 
\sigma^0_{\alpha \dot\alpha} 
+ i {\cal I} \sigma^0_{\alpha \dot\alpha} \Delta_1 \lambda^\alpha
\ee
Thus, writing explicitly $\Pi^i$ and collecting the terms according to the 
order of the derivative of $\cal F$ we have
\bea
i {\cal I} \sigma^0_{\alpha \dot\alpha} \Delta_1 \lambda^\alpha &=& 
-\frac{1}{2} ({\cal F}^{(2)} {\hat v}^{0 i} 
- {\cal F}^{(2) \dagger} {\hat v}^{\dagger 0 i}) 
\eps_1^\alpha \sigma_{i \alpha \dot\alpha}
+ \frac{i}{2} {\cal F}^{(2) \dag}  v^{* 0 i}
\eps_1^\alpha \sigma_{i \alpha \dot\alpha} \nonumber \\
&& + \frac{1}{\sqrt2} {\cal F}^{(3)} \lambda\sigma^{0 i}\psi
\eps_1^\alpha \sigma_{i \alpha \dot\alpha} 
- \frac{1}{\sqrt2} {\cal F}^{(3)} \eps_1 \psi \lambda^\alpha 
\sigma^0_{\alpha \dot\alpha} \nonumber \\
&& - \frac{1}{\sqrt2} {\cal F}^{(3) \dag} 
     \bar\lambda \bar\sigma^{0 i} \bar\psi
\eps_1^\alpha \sigma_{i \alpha \dot\alpha}
- \frac{1}{\sqrt2} {\cal F}^{(3) \dag} 
     \eps_1 \sigma^0 \bar\psi \bar\lambda_{\dot\alpha}
\eea
The terms are arranged such that in the first column there are terms from 
$\Pi^i$ and in the second the other terms. Now we notice that the terms in the 
first line should combine to give the term proportional to $v_{\mu \nu}$
and the other two lines should combine to give the term proportional 
to $D^{\rm on}$ in $\delta_1 \lambda$.  As a matter of fact the first line
gives
\be \label{v}
-\frac{1}{2} ({\cal F}^{(2)} - {\cal F}^{(2) \dag}) {\hat v}^{0 i}
\eps_1^\alpha \sigma_{i \alpha \dot\alpha} 
= -i {\cal I} (\eps_1 \sigma^{\mu \nu} \sigma^0)_{\dot\alpha} v_{\mu \nu}
\ee
and the first term in the second line 
\be \nonumber
\frac{1}{\sqrt2} {\cal F}^{(3)} \lambda^\beta (\sigma^{0 i})_\beta^\gamma
\psi_\gamma \eps_1^\alpha \sigma_{i \alpha \dot\alpha} 
= \frac{1}{2\sqrt2} {\cal F}^{(3)}
(\eps_1 \psi \lambda^\alpha \sigma^0_{\alpha \dot\alpha}
- \eps_1 \lambda \psi^\alpha \sigma^0_{\alpha \dot\alpha})
\ee
combined to the second term in the same line give
\be \label{D1}
- \frac{1}{2\sqrt2} {\cal F}^{(3)}
(\eps_1 \psi \lambda^\alpha \sigma^0_{\alpha \dot\alpha}
+ \eps_1 \lambda \psi^\alpha \sigma^0_{\alpha \dot\alpha})
= \frac{1}{2\sqrt2} {\cal F}^{(3)} \psi \lambda 
\eps_1^\alpha \sigma^0_{\alpha \dot\alpha}
\ee
Finally we can write the first term in the third line as follows
\be \nonumber
\frac{1}{2\sqrt2} {\cal F}^{(3) \dagger} 
(\eps_1 \sigma^0 \bar\psi \bar\lambda_{\dot\alpha}
  - \eps_1 \sigma^0 \bar\lambda \bar\psi_{\dot\alpha})
\ee
which combined with the second term in the same line gives
\be \label{D2}
\frac{1}{2\sqrt2} {\cal F}^{(3) \dag} 
\bar\psi \bar\lambda \eps_1^\alpha \sigma^0_{\alpha \dot\alpha}
\ee
Collecting the terms in (\ref{v}), (\ref{D1}) and (\ref{D2})
we have 
\be
i {\cal I} \sigma^0_{\alpha \dot\alpha} \Delta_1 \lambda^\alpha =
-i {\cal I} (\eps_1 \sigma^{\mu \nu} \sigma^0)_{\dot\alpha} v_{\mu \nu}
+ \frac{1}{2\sqrt2} {\cal F}^{(3)} \psi \lambda 
\eps_1^\alpha \sigma^0_{\alpha \dot\alpha}
+ \frac{1}{2\sqrt2} {{\cal F}^\dagger}''' 
\bar\psi \bar\lambda \eps_1^\alpha \sigma^0_{\alpha \dot\alpha}
\ee
which eventually gives the wanted expression.

\noindent
We want also to show here that the transformations of the dummy fields on-shell
can be obtained by the transformations of the fermions. 
The Euler-Lagrange equations for the fermions, obtained from the Lagrangian 
(\ref{lu1eff}), are given by
\be
{\not\!\bar\partial}^{\dot\alpha \alpha} \psi_{\alpha} = 
\frac{i}{2} f ({\not\!\bar\partial}^{\dot\alpha \alpha}  A) \psi_{\alpha}
- \frac{1}{2} f^{\dagger}
(\frac{1}{\sqrt 2} \bar\sigma^{\mu\nu\!\dot\alpha}_{\quad\dot\beta}
                    \bar\lambda^{\dot\beta} v_{\mu\nu}
+ E \bar\psi^{\dot\alpha}
+ \frac{i}{\sqrt 2} D \bar\lambda^{\dot\alpha})
+ \frac{1}{4} g^{\dagger} \bar\psi^{\dot\alpha} \bar\lambda\bar\lambda
\ee
\be
{\not\!\partial}_{\alpha \dot\alpha} \bar\psi^{\dot\alpha} = 
- \frac{i}{2} f^{\dagger} ({\not\!\partial}_{\alpha \dot\alpha} 
   A^{\dagger}) \bar\psi^{\dot\alpha}
+ \frac{1}{2} f
  (\frac{1}{\sqrt 2} \sigma^{\mu\nu \! \beta}_{\! \alpha}
                    \lambda_{\beta} v_{\mu\nu}
+ E^\dag \psi_{\alpha}
- \frac{i}{\sqrt 2} D \lambda_{\alpha})
- \frac{1}{4} g \psi_{\alpha} \lambda\lambda
\ee
\be
{\not\!\bar\partial}^{\dot\alpha \alpha} \lambda_{\alpha} = 
\frac{i}{2} f ({\not\!\bar\partial}^{\dot\alpha \alpha}  A) \lambda_{\alpha}
- \frac{1}{2} f^{\dagger}
(- \frac{1}{\sqrt 2} \bar\sigma^{\mu\nu\!\dot\alpha}_{\quad\dot\beta}
                    \bar\psi^{\dot\beta} v_{\mu\nu}
+ E^\dag \bar\lambda^{\dot\alpha}
+ \frac{i}{\sqrt 2} D \bar\psi^{\dot\alpha})
+ \frac{1}{4} g^{\dagger} \bar\lambda^{\dot\alpha} \bar\psi\bar\psi
\ee
\be
{\not\!\partial}_{\alpha \dot\alpha} \bar\lambda^{\dot\alpha} = 
- \frac{i}{2} f^{\dagger} ({\not\!\partial}_{\alpha \dot\alpha} 
   A^{\dagger}) \bar\lambda^{\dot\alpha}
+ \frac{1}{2} f
  (-\frac{1}{\sqrt 2} \sigma^{\mu\nu \! \beta}_{\! \alpha}
                    \psi_{\beta} v_{\mu\nu}
+ E \lambda_{\alpha}
- \frac{i}{\sqrt 2} D \psi_{\alpha})
- \frac{1}{4} g \lambda_{\alpha} \psi\psi
\ee
where $f(A,A^{\dagger}) \equiv {\cal F}^{(3)} / {\cal I}$ and 
$g(A,A^{\dagger}) \equiv {\cal F}^{(4)} / {\cal I}$.

\noindent
After a lengthy computation we obtain
\be
\delta_1 E = 0
\ee
\bea
-\frac{i}{\sqrt2} \delta_1 E^\dag &=& 
\eps_1^\alpha {\bigg [} - f^{\dagger}
(\frac{i}{2} (\not\!\partial_{\alpha \dalpha} A^\dag) \bpsi^{\dalpha})
+ \frac{1}{2} \lambda_\alpha [(g-\frac{1}{2i}f^2)\psi\lambda 
- \frac{1}{i4\sqrt2} ff^\dag \bpsi\blambda] \nonumber \\
&+& \frac{1}{2\sqrt2}f
(\sqrt2 \psi_\alpha E^\dag + (\sigma^{\mu\nu}\lambda)_\alpha v_{\mu\nu} 
+ i \lambda_\alpha D) {\bigg ]} 
\eea
and
\bea
-2\sqrt2 \delta_1 D &=& 
\eps_1^\alpha {\bigg [} - f^{\dagger} 
(i\sqrt2 (\not\!\partial_{\alpha \dalpha} A^\dag) \blambda^{\dalpha})
+ \sqrt2 \psi_\alpha [(g-\frac{1}{2i}f^{2})\psi\lambda - 
\frac{1}{2i}f f^{\dagger}\bar\psi\bar\lambda] \nonumber \\
&+& f(\sqrt2 \lambda_\alpha E - (\sigma^{\mu\nu}\psi)_\alpha v_{\mu\nu} 
+ i \psi_\alpha D) {\bigg ]} 
\eea
where we used
\be
\delta f = \delta A (g - \frac{1}{2i} f^{2}) 
+ \delta A^{\dagger} \frac{1}{2i} f f^{\dagger} 
\ee
Comparing these expressions with the Euler-Lagrange equations 
for the fermions we have
\be
\delta_1 E = 0\ , \quad  \delta_1 E^\dag = i \sqrt2 \eps_1\dsl\bpsi\ , \quad 
\delta_1 D = - \eps_1\dsl\blambda
\ee
in agreement with the given Susy transformations.

\section{SU(2) Effective Hamiltonian and Lagrangian}\label{su2comp}

\noindent
In this Appendix we want to show in some details the computations leading
to the Hamiltonian and Lagrangian for the SU(2) sector of the SW model.

\noindent
The Poisson brackets relevant to compute the Hamiltonian are given by 
\bea
\{\eps_1 Q_1 , \bar\eps_1 \bar{Q}_1 \}_{-} &=&
\int d^3 x d^3 y 
\{ \Pi^{a i} \delta_1 v^a_i
+ \delta_1  {\bpsi}^a \pi^a_{\bpsi} 
+ \frac{i}{2}  {\cal F}^{\dagger a b} 
\eps_1 \sigma_i {\blambda}^a v^{* 0 i b} \nonumber \\
&& - \frac{1}{2\sqrt2}  {\cal F}^{\dagger a b c}
\eps_1 \sigma^0 {\bpsi}^a \blambda^b \blambda^c 
+ i  {\cal I}^{a b} \eps_1 \sigma^0 \blambda^b 
\eps^{a c d} A^c A^{d \dagger} \; , \;  \nonumber \\
&& \Pi^{e j} \bar\delta_1 v^e_j
+ \bar\delta_1 A^{e \dagger} \pi^e_{A^\dagger}
+ \frac{1}{\sqrt2} \beps_1 \bpsi^e {\cal F}^{efg \dagger}
(\bpsi^f \bsigma^0 \psi^g + \blambda^f \bsigma^0 \lambda^g) \nonumber \\
&&+ \sqrt2 {\cal I}^{e f} \beps_1 \bsigma^j \sigma^0 \bpsi^f
{\cal D}_j A^e 
+ \frac{i}{2}  {\cal F}^{e f} 
\beps_1 \bsigma_j \lambda^e v^{* 0 j f} \nonumber \\
&&+ \frac{1}{2\sqrt2}  {\cal F}^{efg}
\beps_1 \bsigma^0 \psi^e \lambda^f \lambda^g 
- \beps_1 \pi^e_{\blambda}
\eps^{efg} A^f A^{g \dagger} \}_{-} 
\eea
where $\delta_1 v_i^a = i\eps_1 \sigma_i \blambda^a$, 
$\bar\delta_1 v_j^e = i \beps_1 \bsigma_j \lambda^e$,
$\bar\delta_1 A^{e \dag} = \sqrt2 \beps_1 \bpsi^e$,
$\delta_1 A^b = \sqrt2 \eps_1 \psi^b$,
$\delta_1 \bpsi^a = -i \sqrt2 \eps_1 \not\!{\cal D} A^{a \dag}$ and
$\delta_1 \bar\psi^a \pi^a_{\bar\psi} = 
\delta_1 A^b \pi^b_A + 
\sqrt2 {\cal I}^{ab} \epsilon_1 \sigma^i \bar\sigma^0 \psi^b
{\cal D}_i A^{a \dagger}$.

\noindent
The computation is lengthy but straightforward. There are 44 non-zero 
contributions to $H$ that have to be rearranged and manipulated according to 
the identities above given. Let us write the following useful formulae
\be
\{ \bpsi^a_{\dalpha} \; , \; \psi^b_\alpha \}_{+}
= \{ \blambda^a_{\dalpha} \; , \; \lambda^b_\alpha \}_{+} 
= -i ({\cal I}^{ab})^{-1} \sigma^0_{\alpha \dalpha}
\ee
\bea
\{ \pi^a_A \; , \; \psi^b_\alpha(\lambda^b_\alpha) \}_{-} 
&=& -\frac{i}{2} ({\cal I}^{bc})^{-1} {\cal F}^{cad} 
\psi^d_\alpha(\lambda^d_\alpha) \\
\{ \pi^a_{A^\dag} \; , \; \psi^b_\alpha(\lambda^b_\alpha) \}_{-} 
&=& +\frac{i}{2} ({\cal I}^{bc})^{-1} {\cal F}^{\dag cad} 
\psi^d_\alpha(\lambda^d_\alpha)
\eea
and 
\bea
\{ \eps_1 \sigma_i \blambda^a \; , \; \beps_1 \bsigma_j \lambda^e \}_{-}
&=& i ({\cal I}^{ae})^{-1} \eps_1 \sigma_i \bsigma^0 \sigma_j \beps_1 \\
\{ \eps_1 \sigma_i \blambda^a \; , \; \blambda^f \bsigma^0 \lambda^g \}_{-}
&=& i ({\cal I}^{ag})^{-1} \eps_1 \sigma_i \blambda^f \\
\{ \Pi^{ai} \; , \; {\cal D}_j A^e \}_{-} 
&=&   \eps^{aeh} \delta^i_j A^h \\
\{ \Pi^{ai} (x) \; , \; v^{* 0j f} (y) \}_{-} 
&=& - 2 \eps^{0ijk} (\delta^{af}  \partial^y_k +   \eps^{afh} v^h_k(y))
\delta^{(3)} (\vec{x} - \vec{y})  \\
\{ \eps_1 \sigma_i \blambda^a \; , \; \lambda^f \lambda^g \}_{-}
&=& -i({\cal I}^{af})^{-1} \eps_1 \sigma_i \bsigma^0 \lambda^g 
+ (f \leftrightarrow g) \\
\{ \eps_1 \sigma_i \blambda^a  \; , \; \beps_1 \pi^e_{\blambda} \}_{-}
&=& \delta^{ae} \beps_1 \bsigma_i \eps_1 \\
\{ {\cal D}_\mu A^{\dag a} (x) \; , \; \pi^e_{A^\dag} (y) \}_{-}
&=& (\delta^{ae} \partial^x_i +   \eps^{ade} v_i^d (x)) \delta^{(3)} 
(\vec{x} - \vec{y}) \\
\{ \pi^{a \dalpha}_{\bpsi} \; , \; 
\beps_1 \bpsi^e \bpsi^f \bsigma^0 \psi^g \}_{-} 
&=& \beps_1^{\dalpha} \delta^{ae} \bpsi^f \bsigma^0 \psi^g 
+ \beps_1 \bpsi^e (\bsigma^0 \psi^g)^{\dalpha} \delta^{af} \\
\{ \blambda^b \blambda^c \; , \; \beps_1 \bsigma_j \lambda^e \}_{-}
&=& i({\cal I}^{ec})^{-1} \beps_1 \bsigma_j \sigma^0 \blambda^b 
+ (b \leftrightarrow c) \\
\{ \eps_1 \sigma^0 \bpsi^a \; , \; \bpsi^f \bsigma^0 \psi^g \}_{-}
&=& i ({\cal I}^{ag})^{-1} \eps_1 \sigma^0 \bpsi^f \\
\{ \blambda^b \blambda^c \; , \; \blambda^f \bsigma^0 \lambda^g \}_{-}
&=& i ({\cal I}^{gc})^{-1} \blambda^f \blambda^b +(b \leftrightarrow c) \\
\{ \blambda^b \blambda^c \; , \; \lambda^f \lambda^g \}_{-}
&=&  - i ({\cal I}^{cf})^{-1} \blambda^b \bsigma^0 \lambda^g
- i ({\cal I}^{bg})^{-1} \blambda^c \bsigma^0 \lambda^f \nonumber \\
&& + (f \leftrightarrow g) 
\eea
After commutation we get rid of $\eps_1$ and $\beps_1$ by using 
$\{ \eps_1 Q_1 \; , \; \beps_1 \bar{Q}_1 \}_{-}
= \eps_1^\alpha \beps_1^{\dalpha} 
\{ Q_{1 \alpha} \; , \; \bar{Q}_{1 \dalpha} \}_{+} $ and we trace with 
$\bsigma^{0 \dalpha \alpha}$ obtaining (note that we have still to divide by 
a factor $4i$):

{\bf ``Classical'' terms}

\noindent
{\it Kinetic terms for the e.m. field\footnote{Classical test on 
the e.m. kinetic piece: 
${\cal I}^{-1} \Pi^2 + {\cal I}^{-1} {\cal R} \Pi v^*
+\frac{1}{4} {\cal I}^{-1} ({\cal R}^2 + {\cal I}^2) v^{* 2}
= {\cal I} (E^2 + B^2)$ 
where $v^* = 2 B$ and $\Pi = -({\cal I}E + {\cal R}B)$.}}
\be
-2i ({\cal I}^{ab})^{-1} \Pi^{ai} \Pi^{bi} 
-2i ({\cal I}^{ab})^{-1} {\cal R}^{bc} \Pi^{ai} v^{* 0i c} 
- \frac{i}{2} ({\cal I}^{ae})^{-1} {\cal F}^{\dag ab} {\cal F}^{ef}
v^{* 0i b} v^{* 0i f}
\ee
{\it Kinetic terms for the scalar fields}
\be
-4i ({\cal I}^{ab})^{-1} \pi^a_A (\pi^b_A)^\dag 
+4i {\cal I}^{ab} ({\cal D}^i A^a)({\cal D}^i A^{\dag b})
\ee
{\it Kinetic terms for the spinors}
\be
2i {\cal F}^{ab} \lambda^a \sigma^i {\cal D}_i \blambda^b 
-2i {\cal F}^{\dag ab} \blambda^a \bsigma^i {\cal D}_i \lambda^b 
\ee
and 
\be 
-2 {\cal I}^{ab} \psi^a \sigma^i {\cal D}_i \bpsi^b  
-2 {\cal I}^{ab} \bpsi^a \bsigma^i {\cal D}_i \psi^b  
+ 2 \bpsi^e \bsigma^i \psi^g [\partial_i {\cal I}^{ab} 
+ \eps^{ebc} v_i^b {\cal I}^{gc} +   \eps^{gbc} v_i^b {\cal I}^{ec}] 
\ee
Integrating by parts we have 
\be
- 4 {\cal I}^{ab} (\lambda^a \sigma^i {\cal D}_i \blambda^b
+ \psi^a \sigma^i {\cal D}_i \bpsi^b)
+ 2i (\partial_i {\cal F}^{\dag ab}) \blambda^a \bsigma^i \lambda^b
- \partial_i (2i {\cal F}^{\dag ab} \blambda^a \bsigma^i \lambda^b
+ 2 {\cal I}^{ab} \bpsi^a \bsigma^i \psi^b)
\ee
{\it Yukawa potential}
\bea
&& - i 3 \sqrt2   \eps^{eah} {\cal I}^{ef} A^h \bpsi^f \blambda^a 
-i \sqrt2   \eps^{acd} {\cal I}^{ad} A^c \bpsi^d \blambda^b 
- \frac{3}{\sqrt2}   \eps^{bfg} {\cal F}^{\dag abc} A^f A^{\dag g}
\bpsi^a \blambda^c \nonumber \\
&-& 3 i \sqrt2   \eps^{aeh} {\cal I}^{ab} A^{\dag h} \psi^b \lambda^e 
- i \sqrt2   \eps^{efg} {\cal I}^{ed} A^{\dag g} \psi^f \lambda^d 
-\frac{1}{\sqrt2}   \eps^{acd} {\cal F}^{eag} A^c A^{\dag d}
\psi^e \lambda^g
\eea
By using the properties of $\eps^{abc} {\cal F}^{ade}$ listed in 
Appendix \ref{su2conv} we can recast these terms into
\be
-i 2 \sqrt2   \eps^{abc} {\cal I}^{ad} 
(A^c \bpsi^d \blambda^b + A^{\dag c} \psi^d \lambda^b)
\ee
{\it Higgs potential}
\be
2i \eps^{acd} {\cal I}^{ab} \eps^{bfg} A^c A^{\dag d} A^f A^{\dag g}
\ee

{\bf Purely quantum corrections}

\noindent
{\it Terms that contribute to ${\cal F}^{abc} \sigma_{\mu \nu} v^{\mu \nu}$
and to dummy fields on shell via the two fermions piece $\Pi_{\rm F}$}

\bea
&&  2\sqrt2 ({\cal I}^{ec})^{-1}
{\cal F}^{\dag abc} \Pi^{ei} \bpsi^a \bsigma_{i0} \blambda^b 
- 2\sqrt2 ({\cal I}^{af})^{-1}
{\cal F}^{feg} \Pi^{ai} \psi^e \sigma_{i0} \lambda^g \nonumber \\
&+& \sqrt2 ({\cal I}^{ag})^{-1} {\cal R}^{ab}
{\cal F}^{\dag efg} v^{* 0i b} \bpsi^e \bsigma_{i0} \blambda^f 
- \sqrt2 ({\cal I}^{af})^{-1} {\cal R}^{ab}
{\cal F}^{efg} v^{* 0i b} \psi^e \sigma_{i0} \lambda^g \nonumber \\
&+& i \sqrt2 {\cal F}^{\dag abc} v^{* 0i a} \bpsi^b \bsigma_{i0} \blambda^c 
+ i \sqrt2 {\cal F}^{abc} v^{* 0i a} \psi^b \sigma_{i0} \lambda^c 
\eea
Summing them up we obtain\footnote{
$\Pi + \frac{1}{2}{\cal F}^{(2)} v^{*} = -\frac{1}{2i}{\cal I}\hat{v}^\dag
+ \Pi_{\rm F}$ and 
$\Pi + \frac{1}{2}{{\cal F}^{(2)}}^\dag v^{*} = -\frac{1}{2i}{\cal I}\hat{v}
+ \Pi_{\rm F}$. Also $\hat{v} = E + i B$ and $\hat{v}^\dag = E - i B$.}
\be
-2 \sqrt2 ({\cal I}^{af})^{-1} {\cal F}^{feg} \psi^e\sigma_{i0}\lambda^g 
(\Pi^{ia}+ {\cal F}^{\dag ab} B^{ib}) 
+ 2 \sqrt2 ({\cal I}^{ec})^{-1}{\cal F}^{\dag abc} 
\bpsi^a\bsigma_{i0}\blambda^b (\Pi^{ie}+ {\cal F}^{ed} B^{id}) 
\ee
{\it Terms that contribute to the dummy fields on-shell only}
\bea
&& \frac{i}{4} {\cal F}^{\dag efg} {\cal F}^{cad} ({\cal I}^{gc})^{-1}
\bpsi^e\bpsi^f \psi^a\psi^c  
+ \frac{3i}{4} {\cal F}^{bec} {\cal F}^{efg} ({\cal I}^{ab})^{-1}
\psi^a \psi^c \lambda^f \lambda^g \nonumber \\
&+& \frac{3i}{4} {\cal F}^{\dag bec} {\cal F}^{\dag efg} ({\cal I}^{ab})^{-1}
\bpsi^a \bpsi^c \blambda^f \blambda^g 
+\frac{i}{4} {\cal F}^{\dag abc} {\cal F}^{efg} ({\cal I}^{ae})^{-1}
\blambda^b \blambda^c \lambda^f \lambda^g 
\eea
{\it ${\cal F}^{abcd}$-type of terms}
\be
-\frac{1}{2} ({\cal F}^{abcd} \psi^a \psi^b \lambda^c \lambda^d
- {\cal F}^{\dag abcd} \bpsi^a \bpsi^b \blambda^c \blambda^d)
\ee
Collecting all these terms and dividing by $4i$ we obtain the Hamiltonian 
given in the body of the paper.

{\bf The Lagrangian}

\noindent
The Lagrangian obtained by Legendre transforming the Hamiltonian in 
(\ref{Hsu2}) is given by
\be
{\cal L} \equiv {\cal L}_1 + {\cal L}_2
\ee
where 
\bea
{\cal L}_1 &=&  \frac{1}{2i} {\Big [} 
- \frac{1}{4} {\cal F}^{ab} v^{a \mu \nu} \hat{v}^b_{\mu \nu}
+ \frac{1}{4} {\cal F}^{\dag ab} v^{a \mu \nu} \hat{v}^{\dag b}_{\mu \nu}
{\Big ]} \nonumber \\
&& - {\cal I}^{ab} {\Big [} {\cal D}_\mu A^a {\cal D}^\mu A^{\dag b}
+ i \psi^a \not \!\!{\cal D} \bpsi^b  
+ i \lambda^a \not\!\!{\cal D} \blambda^b \nonumber \\
&& - \frac{1}{\sqrt2} \eps^{adc}  
(A^c \bpsi^b \blambda^d + A^{\dag c} \psi^b \lambda^d) 
+ \frac{1}{2} \eps^{acd}  \eps^{bfg}
A^c A^{\dag d} A^f A^{\dag g} {\Big ]} \nonumber \\
&& - \frac{1}{2} (\partial_\mu {\cal F}^{\dag ab}) 
\blambda^a \bsigma^\mu \lambda^b 
- \frac{1}{2} (\partial_0 {\cal F}^{\dag ab}) 
\bpsi^a \bsigma^0 \psi^b 
\eea
contains the ``classical'' terms, and
\bea
{\cal L}_2 &=& \frac{1}{2i} {\Big [} \frac{1}{\sqrt2} {\cal F}^{abc} 
\lambda^a \sigma^{\mu \nu} \psi^b v^c_{\mu \nu}
- \frac{1}{\sqrt2} {\cal F}^{\dag abc} 
\blambda^a \bsigma^{\mu \nu} \bpsi^b v^c_{\mu \nu} {\Big ]} \nonumber \\
&& +\frac{3}{16} ({\cal I}^{ab})^{-1} {\Big [}
{\cal F}^{acd} {\cal F}^{bef} 
(\psi^d\psi^f \lambda^c\lambda^e - \psi^d\lambda^e \lambda^c\psi^f)
- {\cal F}^{gbc} {\cal F}^{gef} \psi^a\psi^c \lambda^e\lambda^f \nonumber \\ 
&& + {\cal F}^{\dag acd} {\cal F}^{\dag bef}
(\bpsi^d\bpsi^f \blambda^c\blambda^e - \bpsi^d\blambda^e \blambda^c\bpsi^f)
- {\cal F}^{\dag gbc} {\cal F}^{\dag gef}
\bpsi^a\bpsi^c \blambda^e\blambda^f \nonumber \\
&& + {\cal F}^{\dag acd} {\cal F}^{\dag bef}
(\lambda^c\sigma^0\bpsi^f \psi^d\sigma^0\blambda^e 
- \lambda^c\sigma^0\blambda^e \psi^d\sigma^0\bpsi^f ) {\Big ]} \nonumber \\
&& - \frac{1}{16} ({\cal I}^{ab})^{-1} {\cal F}^{\dag acd} {\cal F}^{bef} 
(\bpsi^c\bpsi^d \psi^e\psi^f +  
\blambda^c\blambda^d \lambda^e\lambda^f) \nonumber \\
&& + \frac{1}{2i} 
(\frac{1}{4}{\cal F}^{abcd} \psi^a\psi^b \lambda^c\lambda^d
- \frac{1}{4}{\cal F}^{\dag abcd} \bpsi^a\bpsi^b \blambda^c\blambda^d) 
\label{l4fermi}
\eea
contains the purely quantum terms. 

\noindent
Note that this expressions are in agreement with those obtained by superfield 
expansion \cite{wolf} and correctly reduce to the U(1) Lagrangian given in 
(\ref{lu1eff}) in the limit.

\end{document}